\documentclass[11pt,a4paper]{article}
\usepackage[T1]{fontenc}
\usepackage[utf8]{inputenc}
\usepackage{lmodern}
\usepackage[a4paper,margin=24mm]{geometry}
\usepackage{amsmath,amssymb,bm}
\usepackage{graphicx}
\usepackage{microtype}
\usepackage[numbers,sort&compress]{natbib}

\usepackage[font=small,labelfont=bf,labelsep=period]{caption}
\usepackage{xcolor}
\definecolor{linkblue}{RGB}{25,65,115}
\usepackage[colorlinks=true,linkcolor=linkblue,citecolor=linkblue,urlcolor=linkblue]{hyperref}
\hypersetup{
  pdftitle={Statistical Optical Replication and Mixing of Spatial Information across Wavelengths},
  pdfauthor={Wanting Hou, Wei Liu, Hong-Chao Liu, Zhiyuan Ye, Jun Xiong},
  pdfsubject={Author preprint}
}
\newcommand{\dd}{\mathrm d}
\newcommand{\avg}[1]{\left\langle #1\right\rangle}
\newcommand{\FT}{\mathcal F}
\newcommand{\br}{\bm r}
\newcommand{\bx}{\bm x}
\newcommand{\bq}{\bm q}
\begin{document}
\begin{center}
{\LARGE\bfseries Statistical Optical Replication and Mixing of Spatial Information across Wavelengths\par}
\vspace{0.8em}
{\normalsize Wanting Hou\textsuperscript{1}, Wei Liu\textsuperscript{1}, Hong-Chao Liu\textsuperscript{2},
Zhiyuan Ye\textsuperscript{1,*}, Jun Xiong\textsuperscript{3}\par}
\vspace{0.7em}
{\small
\textsuperscript{1}Faculty of Arts and Sciences, Beijing Normal University, Zhuhai, China\par
\vspace{0.3em}
\textsuperscript{2}Macao Centre for Research and Development in Advanced Materials,
Institute of Applied Physics and Materials Engineering, University of Macau,
Taipa, Macao S.A.R. China\par
\vspace{0.3em}
\textsuperscript{3}School of Physics and Astronomy, Applied Optics Beijing Area Major Laboratory,
Beijing Normal University, Beijing, China\par
\vspace{0.5em}
\textsuperscript{*}Correspondence: Zhiyuan Ye (\href{mailto:yezy@bnu.edu.cn}{\nolinkurl{yezy@bnu.edu.cn}})\par}
\end{center}
\begin{abstract}
Accessing and combining the spatial information carried by light at different wavelengths is important for linking structured-light encoding with spectrally distinct detection channels. Here, we introduce a correlation-enabled linear-optical framework for cross-wavelength imaging, bidirectional replication, and spatial information mixing. Shared random holographic modulation establishes correlations between nondegenerate optical fields, allowing spatial information to be retrieved from joint intensity fluctuations. 
In the demonstrated replicator, a spatial pattern encoded at one wavelength is read out in the detection coordinates of another through intensity correlations. Selecting diffraction-order pairs produces correlation kernels containing either the product or the conjugate product of two input field envelopes. These operations reproduce the spatial algebra underlying sum- and difference-frequency generation, while the optical fields retain their original carrier frequencies. The theory accounts for wavelength-dependent propagation and coordinate scaling, linking the input structures to observable correlation patterns. Together, these results establish a common framework for cross-wavelength spatial information processing through engineered optical correlations and joint detection.
\end{abstract}
\par
\medskip
\noindent{\small\textbf{Abbreviations:} BS, beam splitter;
CMOS, complementary metal--oxide--semiconductor;
DFG, difference-frequency generation;
DP, Dove prism;
GI, ghost imaging;
GS, Gerchberg--Saxton;
ICCD, intensified charge-coupled device;
NPBS, non-polarizing beam splitter;
OAM, orbital angular momentum;
RRH, rotating random hologram;
SFG, sum-frequency generation;
SLM, spatial light modulator.}\par
\medskip

\section{Introduction}\label{sec1}
The transverse structure of light provides a rich space for encoding and manipulating optical information. Beyond their intensity distributions, structured optical fields can carry information in spatial phase and modal composition~\cite{forbes2021structured,dorrah2022tunable}, as illustrated by data transmission using orbital-angular-momentum (OAM) multiplexing~\cite{wang2012terabit}. Many optical tasks also require spatial information to be accessed across wavelengths, because illumination, light–matter interaction, and detection can favor different spectral regions~\cite{barh2019parametric,yuan2023quantum}. 
In addition, the generation of multi-wavelength structured light and the associated wavelength-division multiplexing strategy can greatly increase the channel capacity~\cite{he2022towards,lei2015massive,radwell2016achromatic,zou2022high,wang2023coloured,jang2024wavelength,meng2025ultranarrow,ye2025vortex,ahmed2025customizing,ji2025full,han2026wavelength} in diverse optical information tasks.
Connecting these requirements calls for methods that relate spatial information carried by spectrally distinct fields. Of particular interest is a framework that supports both the retrieval of an existing spatial pattern and the controlled combination of two spatial inputs.

Nonlinear optics provides an established route to transferring and processing structured light across wavelengths~\cite{buono2022nonlinear,singh2026progress}. In second-order three-wave mixing, the source polarizations for sum-frequency generation and difference-frequency generation contain the spatial products \(U_1U_2\) and \(U_1U_2^{*}\), respectively, where \(U_1\) and \(U_2\) denote the input field envelopes. These interactions combine amplitude and phase structures while generating light at a new optical frequency. Experiments have demonstrated coherent frequency conversion of spatial-mode superpositions, as well as mode-dependent control through the structured pump field \cite{li2015sum,steinlechner2016frequency,pereira2017orbital,pinheiro2022observation,xu2023orthogonal,yuan2024high,sanchez2025nonlinear,aguilar2025tailoring,li2025mid}. The resulting spatial conversion response depends on the pump profile, spatial overlap, crystal aperture, and phase-matching acceptance.
By using light to control light, a series of nonlinear optical applications~\cite{qiu2018spiral,hong2020second,singh2024light,chen2025observation} can be realized, including pattern recognition, edge enhancement, and aberration correction.
However, nonlinear optical techniques often require strict phase-matching conditions and intense coherent illumination, and typically suffer from relatively low conversion efficiencies, which greatly limits their practical applicability.

Correlation optics offers a linear route in which spatial information is accessed and transferred through joint measurements. Two-photon correlated imaging, also known as Ghost imaging (GI), was first established using entangled photon pairs~\cite{pittman1995optical}, and such imaging protocols, owing to spontaneous parametric down-conversion in nonlinear crystals, can be readily implemented between nondegenerate photon pairs~\cite{chan2009two,aspden2015photon,ma2025quantum}; consequently, the near-infrared image of an object can be transferred to visible photons that do not interact with the object at all. Subsequently, GI was demonstrated with classically correlated thermal or pseudothermal light~\cite{ferri2005high}. However, cross-wavelength thermal-light GI still relies heavily on various nonlinear optical techniques~\cite{yu2016ghost,shu2026seeing}, including four-wave mixing, because nondegenerate thermal light fields naturally decorrelate as the frequency difference increases~\cite{deng2018single,ye2020pseudo,xiong2025polychromatic}. Therefore, finding classical linear alternatives—neither quantum nor nonlinear—for coherent optical information interaction and processing across wavelengths has become of considerable practical value, especially when nonlinear optical techniques are costly or practically infeasible.
Some broadband and achromatic optical modulation techniques are key to establishing nondegenerate spatial correlations. Radwell \textit{et al.} demonstrated broadband vector-vortex generation using a glass cone that combines wavelength-independent geometric phase with weakly dispersive total-internal-reflection retardance~\cite{radwell2016achromatic}.
Recent work has shown that shared random holographic modulation~\cite{ye2025random,hou2025manipulating,ye2026all} makes it possible to establish all-purpose spatial correlations in pseudo-thermal light over a broad bandwidth (from ultraviolet to near-infrared), in which two-color sub-wavelength interference was well demonstrated with sum- and difference-type spatial fringe relations. These advances establish the physical resources for a broader processing framework. 

Here, using random holography as an intermediary, we introduce a correlation-enabled statistical optical framework for cross-wavelength correlation imaging, optical replication, spatial information mixing, and statistical optical-field manipulation. 
Within a shared, wavelength-independent random-modulation model, selecting the diffraction-order pair gives access to correlation kernels containing either \(U_1U_2\) or \(U_1U_2^*\). Their propagation establishes a correspondence with the spatial source terms of sum- and difference-frequency generation, while the measured outputs are obtained from joint intensity fluctuations. 
Building on broadband spatial correlations, we demonstrate a common platform for cross-wavelength imaging, bidirectional pattern replication, and two-input spatial mixing, and develop a conditional-transmission extension for preparing structured statistical output fields.
This statistical framework could serve as an information relay or a correlation-based spatial-field multiplier in multi-wavelength structured-light communication, offering a potential route to low-cost readout of spatial field products and statistical beam shaping in less accessible spectral regions~\cite{ceddia2022ghost,ceddia2023universal,kingston2025neutron}.





 \section{Theory}\label{sec2}

\begin{figure}[htbp]%
\centering
\includegraphics[width=0.96\linewidth]{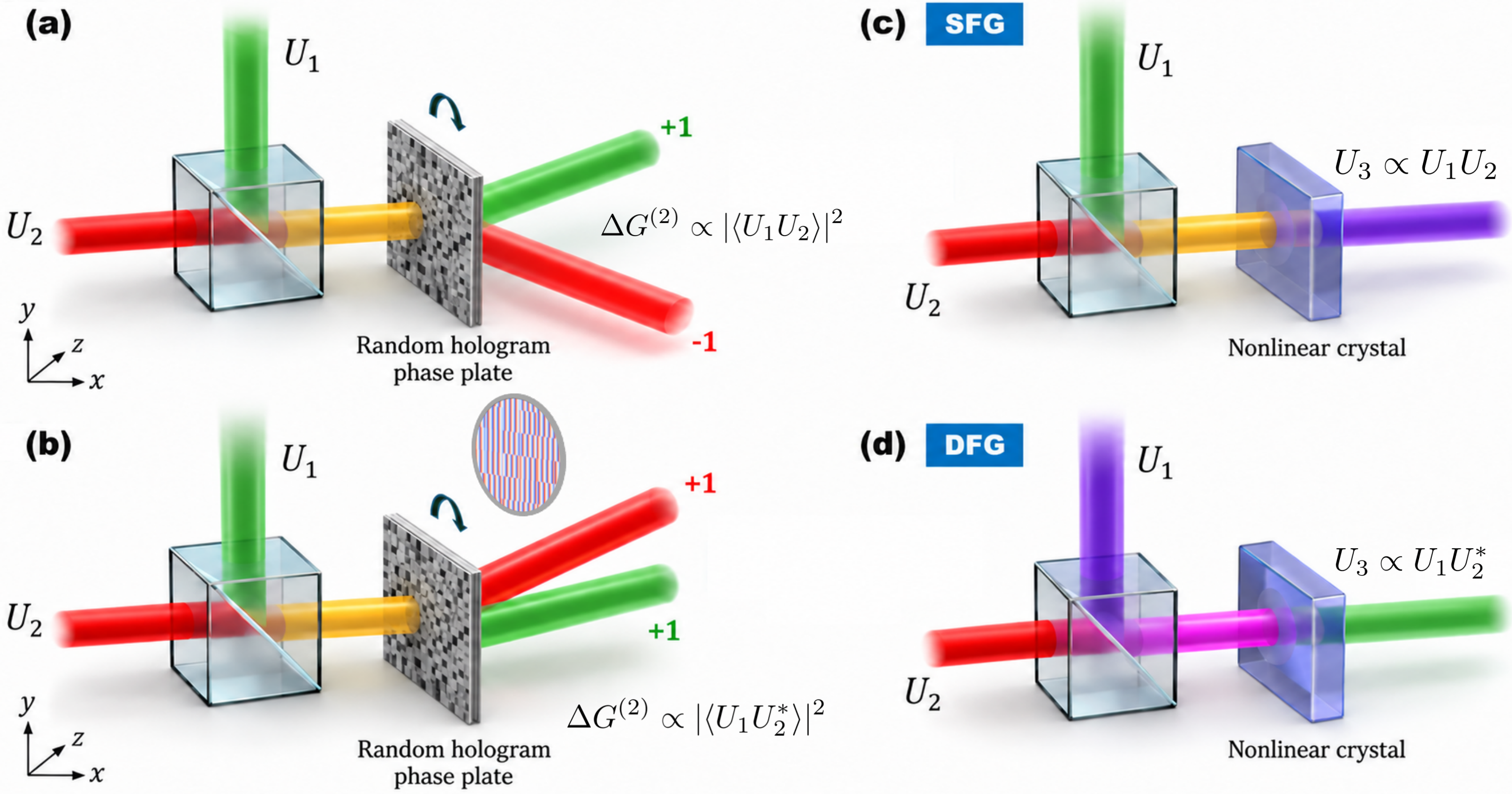}
\caption{Statistical counterparts of spatial operations in nonlinear optics. (a) Opposite diffraction orders carry conjugate random modulations and support a correlation kernel containing the product of the input envelopes \(\langle U_1U_2 \rangle\). (b) Same diffraction orders carry identical random modulations and support their conjugate product \(\langle U_1U^*_2 \rangle\). (c,d) Corresponding spatial products in sum- and difference-frequency generation. The statistical schemes preserve the incident optical frequencies and require joint detection or conditionally triggered measurement, whereas the nonlinear schemes generate a field at a new frequency. 
}\label{fig1}
\end{figure}

As shown in Fig.~\ref{fig1}, the central idea of this work is to replace the nonlinear crystal with a rotating random holographic phase plate, using shared statistical fluctuations to mediate the processing of coherent spatial information across wavelengths. The plate comprises micro-gratings with a common period and randomized orientations and displacement phases~\cite{ye2025random}. Their \(\pm1\) diffraction orders form annular angular spectra, while rotation samples an ensemble of random realizations shared by the two incident wavelengths. 
As shown in Figs.~\ref{fig1}(a)(b), selecting opposite or equal diffraction orders yields conjugate or identical random modulations, respectively, establishing correlation kernels that contain nonconjugated or conjugated products of the incident spatial fields. These products provide statistical counterparts to the spatial operations underlying sum- and difference-frequency generation in Figs.~\ref{fig1}(c)(d), with the information accessed through intensity correlations and the optical carrier frequencies remaining unchanged.

\begin{figure}[htbp]%
\centering
\includegraphics[width=0.98\linewidth]{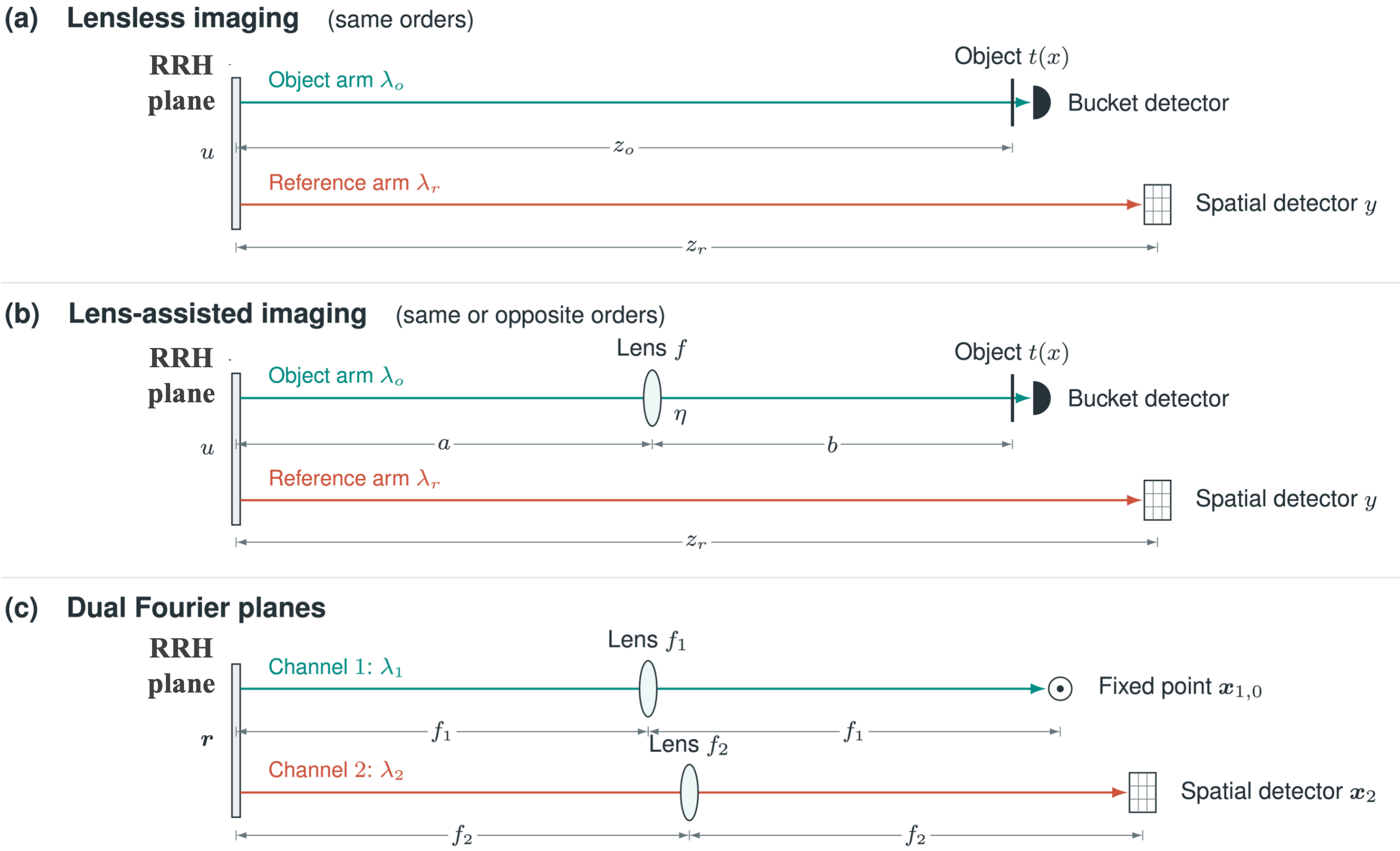}
\caption{Geometries and coordinates used in the theory. The selected two-color channels are unfolded onto separate forward-propagating axes. (a) Lensless nondegenerate correlation imaging using the same order diffracted beams; (b) Lens-assisted nondegenerate correlation imaging; (c) Fourier readout for replication and mixing: the source and detector are the front and back focal planes of each lens. In (a)(b), the intensity correlation is established between a spatially unresolved bucket detection signal and an array detector, whereas in (c) it is established between a fixed point detector and an array detector.
}\label{fig2}
\end{figure}

We develop the theory through three related configurations. The three functions considered below differ in where the information is introduced and how the correlations are read.
First, in nondegenerate correlation imaging, an object modulates one selected diffracted beam, and its image is reconstructed in the detection coordinates of the other wavelength through joint intensity measurements. Second, in optical replication, the information is instead carried by an incident structured beam of $\lambda_1$, and a smooth reference beam at $\lambda_2$ enables this information to be recovered in the cross-correlation function.
Third, when both incident beams are structured, their complex spatial envelopes enter the correlation kernel through multiplication or conjugate multiplication; the system thereby operates as a spatial mode mixer, allowing the coupled spatial information of the two colors to be read out within the same statistical framework.
Together, these operations define an approach to cross-wavelength spatial information processing in which linear optical propagation supplies the fields and joint intensity measurements supply the computational readout.
We first define the shared source and measured covariance, then derive these three functions using the geometries in Fig.~\ref{fig2}.

\subsection{Shared source and correlation readout}\label{sec:source}

Let $U_j(\br)$ be the prescribed complex envelope of input $j$ at wavelength $\lambda_j$, where $\br$ is a two-dimensional coordinate in the hologram plane. Lowercase $e_{j,0}$ and $e_j$ denote the random envelopes immediately after channel selection and at detection, respectively. All fields are scalar and quasi-monochromatic. After registering the channel coordinates and removing deterministic carrier tilts, the ideal source model is
\begin{equation}
\begin{aligned}
e_{1,0}(\br)&=c_1U_1(\br)\xi(\br),\\
e_{2,0}^{\rm same}(\br)&=c_{2,\rm same}U_2(\br)\xi(\br),\\
e_{2,0}^{\rm opp}(\br)&=c_{2,\rm opp}U_2(\br)\xi^*(\br).
\end{aligned}
\label{eq:source}
\end{equation}
Here $\xi$ is a dimensionless random modulation shared by the two colors, and the coefficients $c_j$ describe deterministic diffraction amplitudes and phases. The labels `\emph{same}' and `\emph{opp}' identify same-order and opposite-order diffracted beams. In the opposite-order channel, conjugation acts only on the random modulation, not on the prescribed input $U_2$. 

The source statistics are specified by
\begin{equation}
\begin{aligned}
\avg{\xi(\br)}=0,\,\,\avg{\xi(\br)\xi(\br')}=0,\,\,
J(\br,\br')=\avg{\xi(\br)\xi^*(\br')},
\end{aligned}
\label{eq:sourcecoherence}
\end{equation}
where the function $J$ describes the spatial extent and coherence of the random source. Angle brackets denote averaging over hologram realizations for prescribed input envelopes, with the independent global carrier phases factored out. This convention preserves the spatial phase information relevant to the calculation, and the measured intensities are invariant under independent global phase offsets of either input.

Propagation from the rotating random hologram (RRH) plane to either observation plane is linear:
\begin{equation}
e_j(\bx_j)=\int h_j(\bx_j,\br)e_{j,0}(\br)\,\dd^2\br,
\label{eq:propagation}
\end{equation}
where the impulse response $h_j$ is the complex field at detector coordinate $\bx_j$ produced by a unit point excitation at $\br$. It includes propagation, lenses, apertures, and the wavelength $\lambda_j$. 

The experimentally accessible quantity is the intensity covariance, also referred to as the connected intensity correlation function
\begin{equation}
\begin{aligned}
\Delta G^{(2)}(\bx_1,\bx_2)=\avg{\Delta I_1(\bx_1)\Delta I_2(\bx_2)},
\end{aligned}
\label{eq:covariance}
\end{equation}
with $I_j=|e_j|^2$ and $\Delta I_j=I_j-\avg{I_j}$. The raw correlation is $G^{(2)}=\avg{I_1I_2}=\avg{I_1}\avg{I_2}+\Delta G^{(2)}$, and the normalized correlation function is $g^{(2)}=1+\Delta G^{(2)}/\avg{I_1}\avg{I_2}$.
For zero-mean jointly Gaussian detected envelopes, fourth-moment factorization gives~\cite{picinbono1997second,cui2026correlation}
\begin{equation}
\Delta G^{(2)}=|\avg{e_1e_2^*}|^2+|\avg{e_1e_2}|^2.
\label{eq:siegert}
\end{equation}
Equation~\eqref{eq:siegert} mathematically explains why intensity correlation measurements (i.e. intensity covariance) can directly retrieve information about the first-order coherence of the field without the aid of an amplitude interferometer. 
Note that previous theoretical descriptions under Gaussian statistics considered only joint circularity (corresponding to the first correlation term in Eq.~\eqref{eq:siegert}), whereas discussions of correlations between opposite diffraction orders must take joint non-circularity~\cite{picinbono1994circularity} into account, corresponding to the second correlation term in Eq.~\eqref{eq:siegert}.
For the ideal source in Eq.~\eqref{eq:source}, one of the two terms in Eq.~\eqref{eq:siegert} vanishes. It is therefore useful to define
\begin{equation}
\begin{aligned}
\Gamma_{\rm same}&=\avg{e_1(e_2^{\rm same})^*}, \,\, \Gamma_{\rm opp}=\avg{e_1e_2^{\rm opp}},\\
C_\beta&=|\Gamma_\beta|^2,
\quad \beta\in\{\mathrm{same},\mathrm{opp}\},
\end{aligned}
\label{eq:kernels}
\end{equation}
where $\Gamma$ is a complex correlation amplitude, whereas $C_{\beta}$ is the real, background-subtracted intensity measurement. $C_{\beta}$ represents the intensity covariance between the two detection positions, quantifying their correlated intensity fluctuations for the selected diffraction-order configuration \(\beta\).

Substituting the source fields and propagators yields the general two-color correlation kernels:
\begin{equation}
\begin{aligned}
\Gamma_{\rm same}(\bx_1,\bx_2)
&=A_{\rm same}\iint J(\br,\br')U_1(\br)U_2^*(\br')\\[-2pt]
&\quad\times h_1(\bx_1,\br)h_2^*(\bx_2,\br')\,\dd^2\br\,\dd^2\br',\\
\Gamma_{\rm opp}(\bx_1,\bx_2)
&=A_{\rm opp}\iint J(\br,\br')U_1(\br)U_2(\br')\\[-2pt]
&\quad\times h_1(\bx_1,\br)h_2(\bx_2,\br')\,\dd^2\br\,\dd^2\br',
\end{aligned}
\label{eq:finitekernel}
\end{equation}
where $A_{\rm same}=c_1c_{2,\rm same}^*$ and $A_{\rm opp}=c_1c_{2,\rm opp}$. 

To expose the spatial operation, consider a source coherence length much shorter than the resolved spatial variations. The approximation $J(\br,\br')\simeq S(\br)\delta^{(2)}(\br-\br')$ introduces a nonnegative correlation weight $S$ that includes the illuminated aperture and effective coherence area. Define
\begin{equation}
\mathcal B_{\rm same}=S U_1U_2^*,\qquad
\mathcal B_{\rm opp}=S U_1U_2.
\label{eq:products}
\end{equation}
The two kernels reduce to
\begin{equation}
\begin{aligned}
\Gamma_{\rm same}&\simeq A_{\rm same}\int
\mathcal B_{\rm same}(\br)h_1(\bx_1,\br)h_2^*(\bx_2,\br)\,\dd^2\br,\\
\Gamma_{\rm opp}&\simeq A_{\rm opp}\int
\mathcal B_{\rm opp}(\br)h_1(\bx_1,\br)h_2(\bx_2,\br)\,\dd^2\br.
\end{aligned}
\label{eq:bilinear}
\end{equation}
These equations separate the local spatial product $\mathcal B$ from its wavelength-dependent propagation. They provide the common starting point for the three readouts below. 

\subsection{Nondegenerate correlation imaging}\label{sec:imaging}

For imaging, we relabel the two channels by their roles: $o$ denotes the object arm and $r$ the spatially resolved reference arm shown in Fig.~\ref{fig2}(a). These labels may be assigned to either input wavelength. We use one transverse dimension to display the geometry, with source coordinate $u$, object coordinate $x$, and reference-detector coordinate $y$. The incident fields have approximately planar wavefronts, with negligible residual phase curvature, and their nonnegative amplitude weighting is absorbed into $S(u)$.

An object of amplitude transmission $t(x)$ is placed immediately before a bucket detector with collection weight $w(x)$. The bucket signal and reconstructed image are
\begin{equation}
\begin{aligned}
&B=\int w(x)|t(x)|^2 I_o(x)\,\dd x,\\
&\mathcal G_\beta(y)=\avg{\Delta B\,\Delta I_r(y)}=\int w(x)|t(x)|^2 C_\beta(x,y)\,\dd x.
\end{aligned}
\label{eq:bucket}
\end{equation}
The image therefore probes intensity transmission, with the point-to-point covariance serving as its spatial response. Complete collection of the transmitted intensity does not, by itself, recover an arbitrary object phase.

\subsubsection{Lensless geometry}

We first consider the geometry of lensless correlation imaging~\cite{cao2005geometrical} as shown in Fig.~\ref{fig2}(a). The one-dimensional Fresnel response is $h_j(x,u)\propto\exp[\mathrm i\pi(x-u)^2/(\lambda_j z_j)]$. For same-order channels, conjugation of the reference propagator gives
\begin{equation}
\begin{aligned}
\Gamma_{\rm same}(x,y)\propto\int S(u)
&\exp\!\left[\mathrm i\pi\!\left(\frac{1}{\lambda_o z_o}-\frac{1}{\lambda_r z_r}\right)u^2\right]\\
{}\times&\exp\!\left[-\mathrm i2\pi\!\left(\frac{x}{\lambda_o z_o}-\frac{y}{\lambda_r z_r}\right)u\right]\dd u,
\end{aligned}
\label{eq:lensless}
\end{equation}
up to factors and output phases independent of $u$. The quadratic source phase cancels when
\begin{equation}
\lambda_o z_o=\lambda_r z_r\equiv\Lambda.
\label{eq:match}
\end{equation}
Here $\Lambda$ has dimensions of length squared. This two-color matching condition~\cite{chan2009two} generally requires different physical path lengths. At the matched distances,
\begin{equation}
\begin{aligned}
C_{\rm same}(x,y)&\propto p(x-y),\\
p(v)&=\left|\int S(u)\mathrm{e}^{-\mathrm i2\pi vu/\Lambda}\,\dd u\right|^2.
\end{aligned}
\label{eq:lenslessimage}
\end{equation}
Consequently, Eq.~\eqref{eq:bucket} gives a unit-magnification upright image blurred by $p$.
For instance, a uniform source weight of width $D$ gives $p(v)\propto\operatorname{sinc}^2(Dv/\Lambda)$, where $\operatorname{sinc}(q)=\sin(\pi q)/(\pi q)$.  Note that the wavelength ratio does not introduce an additional image-scaling factor. For opposite-order channels, the two differences in parentheses in Eq.~\eqref{eq:lensless} become sums. Positive free-space distances cannot then satisfy this quadratic-phase-cancellation condition in the uncompensated geometry. Therefore, the nondegenerate lensless imaging configuration can only be implemented through intensity correlations between the same diffraction orders.

\subsubsection{Lens-assisted geometry}

Figure~\ref{fig2}(b) adds a converging lens of focal length $f>0$ at $\lambda_o$ in the object arm. The physical distances are $a$ from source to lens, $b$ from lens to object, and $z_r$ from source to reference detector. The lens coordinate is $\eta$, and its pupil is $P(\eta)$. With a broad, nearly uniform source, composition of the propagators gives the signed effective distances
\begin{equation}
\begin{aligned}
d_{\rm same}=a-\rho z_r,\,\,\,\,d_{\rm opp}=a+\rho z_r,
\end{aligned}
\label{eq:effective}
\end{equation}
with $\rho=\lambda_r/\lambda_o$.
These are distances in the correlation calculation, not additional physical propagation paths. For either branch and $d_\beta\ne0$, one obtains
\begin{equation}
\begin{aligned}
\Gamma_\beta(x,y)\propto\int P(\eta)\exp\!\bigg\{
\frac{\mathrm i\pi}{\lambda_o}\bigg[
\frac{(x-\eta)^2}{b}-\frac{\eta^2}{f}
+\frac{(\eta-y)^2}{d_\beta}\bigg]\bigg\}\,\dd\eta.
\end{aligned}
\label{eq:lenskernel}
\end{equation}
The opposite signs in Eq.~\eqref{eq:effective} follow from conjugating, or not conjugating, the reference propagator. All optical beams still propagate forward.
The coefficient of $\eta^2$ vanishes at the imaging condition
\begin{equation}
\frac{1}{d_\beta}+\frac{1}{b}=\frac{1}{f}.
\label{eq:lenscondition}
\end{equation}
The remaining pupil transform is centered where $x/b+y/d_\beta=0$. The magnification from object coordinate to reconstructed reference coordinate is therefore
\begin{equation}
m_\beta\equiv\frac{y}{x}=-\frac{d_\beta}{b}
=-\frac{f}{b-f}.
\label{eq:magnification}
\end{equation}
For a narrow point-spread function, we have $\mathcal G_\beta(y)\propto|t(y/m_\beta)|^2$. 
At their respective imaging conditions, the same- and opposite-order channels have transverse magnifications \(m_{\mathrm{same}}=-(a-\rho z_r)/b\) and \(m_{\mathrm{opp}}=-(a+\rho z_r)/b\), respectively. 
In this configuration, one has $|m_{opp}|>|m_{same}|$.
The magnification is closely related to the wavelength ratio $\rho$. The sign of the magnification determines whether the observed correlation image is upright or inverted.


\subsection{Replication of an incident spatial pattern}\label{sec:replication}

In the two-color replicator, a pattern is encoded in $U_1$ before entering the RRH plane. 
A fixed point at $\bx_{1,0}$ in the first detector plane is correlated with the spatially resolved second arm, $C_\beta(\bx_{1,0},\bx_2)=|\Gamma_\beta(\bx_{1,0},\bx_2)|^2$.
This is a two-arm reconstruction displayed in the $\lambda_2$ detector coordinates. The second-arm mean intensity alone does not acquire the structure of an independently varied $U_1$.

The two Fourier systems in Fig.~\ref{fig2}(c) provide a concrete mapping. Each RRH plane is a front focal plane and each detector a back focal plane. Define the transverse spatial frequency and Fourier transform by
\begin{equation}
\FT[F](\bq)=\int F(\br)\mathrm{e}^{-\mathrm i\bq\cdot\br}\,\dd^2\br \qquad
\bq_j=\frac{2\pi\bx_j}{\lambda_jf_j}.
\label{eq:fouriercoordinates}
\end{equation}
Within the transmitted spatial bandwidth, Eq.~\eqref{eq:bilinear} becomes
\begin{equation}
\begin{aligned}
\Gamma_{\rm same}(\bq_1,\bq_2)&\propto\FT[\mathcal B_{\rm same}](\bq_1-\bq_2),\\
\Gamma_{\rm opp}(\bq_1,\bq_2)&\propto\FT[\mathcal B_{\rm opp}](\bq_1+\bq_2).
\end{aligned}
\label{eq:fourier}
\end{equation}
The conjugation that distinguishes the input products also determines the sign of the detected spatial-frequency coordinate.
For a uniform second input $U_2=U_0$, nearly uniform source weighting over $U_1$, and an on-axis reference point $\bq_1= \mathbf{0}$, both branches reproduce the Fourier intensity associated with the first input. Let $I_{F}(\bx)\propto|\FT[U_1](2\pi\bx/(\lambda_1f_1))|^2$ denote that intensity without random modulation. Then
\begin{equation}
\begin{aligned}
\Delta G^{(2)}_{\rm same}({\bf0}, \bx_2)\propto I_{F}(-\kappa\bx_2),\,\,\,\, \Delta G^{(2)}_{\rm opp}( {\bf 0}, \bx_2)\propto I_{F}(+\kappa\bx_2),
\end{aligned}
\label{eq:replicamap}
\end{equation}
with $\kappa={\lambda_1f_1}/{\lambda_2f_2}$.
To reconstruct a target intensity $|A_{\rm tar}(\bq)|^2$ in this geometry, one may encode $S U_1\propto\FT^{-1}[A_{\rm tar}]$, subject to the available aperture and input modulation. 
Compared to correlation imaging requiring bucket detection, a point detector here selects a single spatial sample. An extended detection area instead integrates $C_\beta$ over that area and can blur the reconstructed pattern. 


\subsection{Mixing of two-color structured inputs}\label{sec:mixing}

When both incident fields are structured, the reference is no longer spatially uniform and the same readout becomes a spatial mixer. Substituting the products in Eq.~\eqref{eq:products} into Eq.~\eqref{eq:fourier} gives the measured mixing signals explicitly:
\begin{equation}
\begin{aligned}
\Delta G^{(2)}_{\rm same}(\bx_1,\bx_2)
&\propto\left|\FT[S U_1U_2^*](\bq_1-\bq_2)\right|^2,\\
\Delta G^{(2)}_{\rm opp}(\bx_1,\bx_2)
&\propto\left|\FT[S U_1U_2](\bq_1+\bq_2)\right|^2.
\end{aligned}
\label{eq:mixing-connected}
\end{equation}
Setting $\bx_1=\bx_{1,0}$ recovers the point-reference readout in Fig.~\ref{fig2}(c). Writing $U_j(\br)=A_j(\br)e^{\mathrm i\phi_j(\br)}$, the local products are
\begin{equation}
\begin{aligned}
\mathcal B_{\rm same}=S A_1A_2 \mathrm{e}^{\mathrm i(\phi_1-\phi_2)},\,\,\,
\mathcal B_{\rm opp}=S A_1A_2 \mathrm{e}^{\mathrm i(\phi_1+\phi_2)}.
\end{aligned}
\label{eq:phasearithmetic}
\end{equation}
Same-order selection thus subtracts the input spatial phases, whereas opposite-order selection adds them. The multiplication occurs in the source-plane correlation kernel, and $G^{(2)}_\beta$ measures the squared modulus of its propagated Fourier transform.

This correspondence can be stated directly in terms of difference-frequency generation (DFG) and sum-frequency generation (SFG) shown in Figs.~\ref{fig1}(c)(d). For input angular frequencies $\omega_j=2\pi c/\lambda_j$, where $c$ is the speed of light in vacuum, the local second-order source polarizations contain
\begin{equation}
\begin{aligned}
\mathcal P_{\rm DFG}^{(2)}(\br)\propto\chi^{(2)} U_1U^*_2,\,\,\,\,
\mathcal P_{\rm SFG}^{(2)}(\br)\propto\chi^{(2)} U_1U_2.
\end{aligned}
\label{eq:nonlinear}
\end{equation}
The appropriate effective susceptibility is understood for each process. The generated frequencies are $\omega_{\rm DFG}=\omega_1-\omega_2$ and $\omega_{\rm SFG}=\omega_1+\omega_2$, with $\omega_1>\omega_2$ assumed for the displayed positive-frequency DFG component. Thus the same-order product $\mathcal B_{\rm same}$ is the spatial counterpart of the DFG source term, while the opposite-order product $\mathcal B_{\rm opp}$ is the counterpart of the SFG source term.

For weak conversion in a thin crystal, assume negligible input depletion, diffraction, and transverse walk-off within the interaction region, together with an approximately uniform phase-matching response over the admitted spatial bandwidth. The generated exit-plane envelopes are then proportional to these input products. Their Fourier-plane intensities therefore satisfy
\begin{equation}
\begin{aligned}
I_{{\rm DFG}}(\bq)&\propto\left|\FT[ U_1U_2^* ](\bq)\right|^2,\\
I_{{\rm SFG}}(\bq)&\propto\left|\FT[ U_1U_2 ](\bq)\right|^2.
\end{aligned}
\label{eq:nonlinear-fourier}
\end{equation}
Here $\bq$ is the transverse spatial frequency in the nonlinear Fourier plane, calibrated using the generated wavelength and the corresponding Fourier-lens focal length. For approximately uniform $S$ and matched effective apertures, Eq.~\eqref{eq:mixing-connected} consequently gives the explicit readout correspondence
\begin{equation}
\begin{aligned}
\Delta G^{(2)}_{\rm same}(\bx_1,\bx_2)
&\propto I_{{\rm DFG}}(\bq_1-\bq_2),\\
\Delta G^{(2)}_{\rm opp}(\bx_1,\bx_2)
&\propto I_{{\rm SFG}}(\bq_1+\bq_2).
\end{aligned}
\label{eq:mixing-correspondence}
\end{equation}
This relation compares spatial distributions after their wavelength-dependent Fourier coordinates have been registered; it does not equate the signal strengths or conversion efficiencies. In particular, an on-axis reference point gives the DFG-like distribution at $-\bq_2$ and the SFG-like distribution at $+\bq_2$. 
The statistical scheme leaves both optical carriers unchanged and retrieves the processed information from connected intensity correlations. It therefore reproduces the spatial multiplication and phase addition or subtraction associated with DFG and SFG, without generating a frequency-converted coherent output field.

An illustrative case is the mixing of azimuthal modes~\cite{yao2011orbital}, $U_j(r,\varphi)=a_j(r)\mathrm{e}^{\mathrm i\ell_j\varphi}$, with $\ell$ being the integer topological charge of the helical phase. 
For rotationally symmetric source weighting and apertures, Eq.~\eqref{eq:phasearithmetic} gives
\begin{equation}
\ell_{\rm same}=\ell_1-\ell_2,\qquad
\ell_{\rm opp}=\ell_1+\ell_2.
\label{eq:oam}
\end{equation}
These indices describe the angular phase of the correlation kernel and follow the corresponding DFG and SFG phase arithmetic. 

\section{Results}\label{sec3}

\subsection{Nondegenerate correlation imaging}\label{sec:results-imaging}

We first examine the imaging configuration of Section~\ref{sec:imaging}, in which the object is placed in one of the selected diffracted beams downstream of the RRH. 
A He--Ne laser at $632.8~\mathrm{nm}$ and a semiconductor laser at $532~\mathrm{nm}$ are expanded, collimated, and combined on the RRH. 
The RRH was fabricated on a quartz substrate following Ref.~\cite{ye2025random}, with a grating period of
$38.6~\mu\mathrm{m}$ and a subgrating size of $60~\mu\mathrm{m}$.
Its rotation produces successive realizations of the shared random modulation described in Eq.~\eqref{eq:source}. The annular diffraction patterns and selected opposite regions are shown in the inset of Fig.~\ref{fig:results-lensless}(a). Spatial selection, beam splitting, and narrow-band filters provide either two copies of the same diffraction order or a pair of opposite orders, with independently selected detection wavelengths.
An amplitude mask (O) bearing the letters ``GI'' is placed immediately before the CMOS-based detector (Daheng Imaging, MER-231-41U3M, the pixel pitch is $5.86~\mu\mathrm m$) in the object arm. Its transmitted intensity is integrated over the selected detector region to form a scalar bucket signal, whereas the CMOS-based detector retains spatial resolution in the reference arm. The resulting bucket--reference correlation is the readout described by Eq.~\eqref{eq:bucket}. In the figures, the localized peaks in the normalized point-to-point correlation $g^{(2)}$ characterize the spatial correlation response, and the adjacent reconstructed images show the object information recovered by bucket--reference correlation. The reconstruction is governed by the object-weighted integral of the point-to-point covariance, rather than by the point-to-point correlation map alone. 

\subsubsection{Lensless imaging across wavelengths}

\begin{figure}[htbp]
\centering
\includegraphics[width=0.88\linewidth]{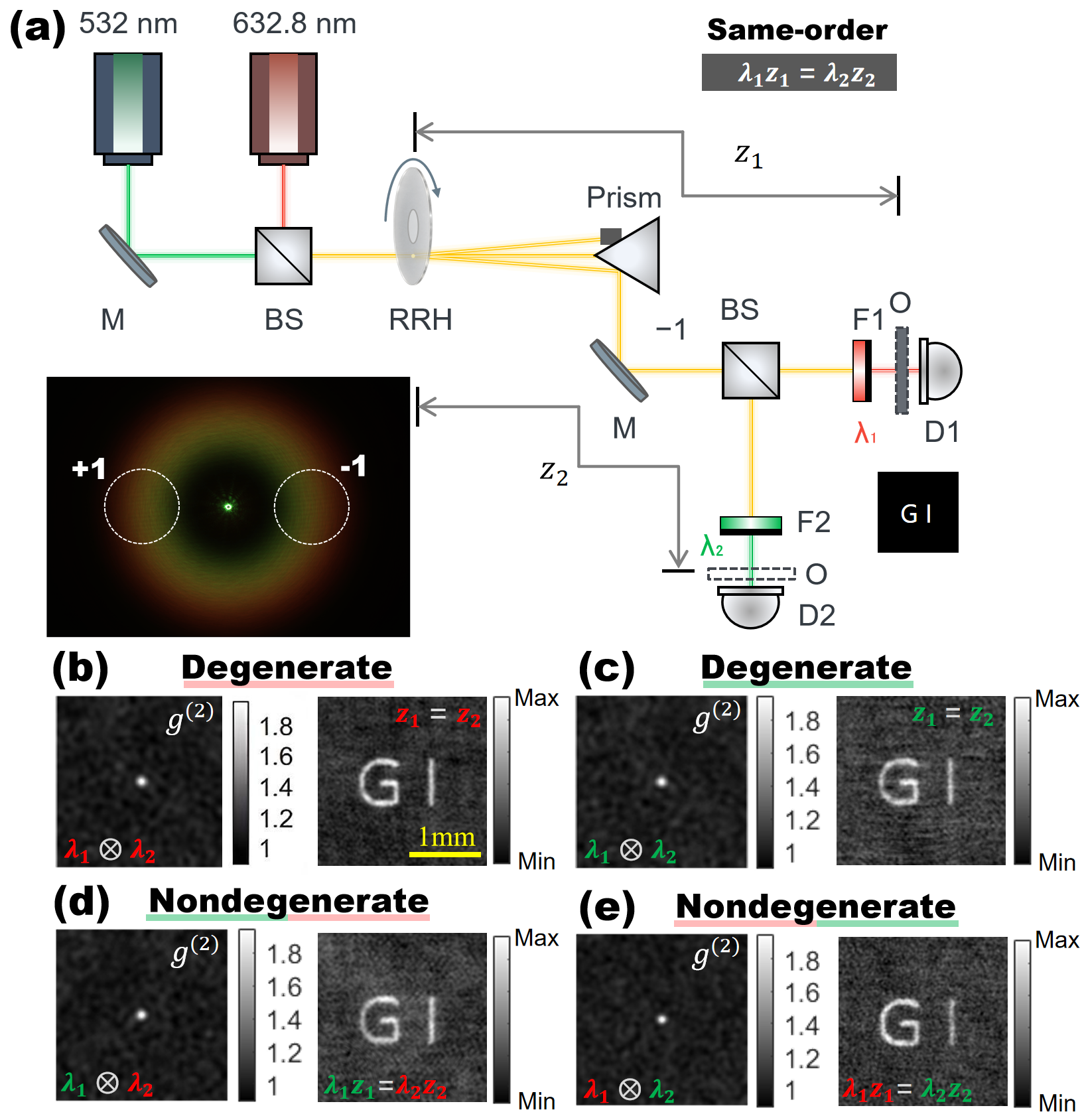}
\caption{Lensless correlation imaging with same-order channels. (a) Experimental arrangement using $532$ and $632.8~\mathrm{nm}$ illumination of the RRH. A selected $-1$ diffraction region is split into two detection arms, with two narrowband spectral filters selecting the detection wavelengths. The two dashed object positions indicate alternative bucket-arm assignments. The inset shows the annular diffraction pattern and the selected opposite regions. (b,c) Degenerate red--red and green--green measurements. (d,e) Nondegenerate measurements with green-object/red-reference and red-object/green-reference assignments, respectively. Each pair contains a normalized point-to-point correlation map $g^{(2)}$ and the corresponding bucket--reference image. The wavelength labels within each panel identify that measurement configuration. 
M, mirror; BS, beam splitter; F1 and F2, spectral filters; O, object; D1 and D2, detectors.
$2\times10^4$ realizations contribute to the image quality.
}
\label{fig:results-lensless}
\end{figure}

Figure~\ref{fig:results-lensless}(a) shows the same-order arrangement corresponding to Fig.~\ref{fig2}(a). The selected $-1$ diffracted beam is divided into object and reference arms, and filters select the detected colors. With the same wavelength in both arms, equal propagation distances recover the degenerate case of Eq.~\eqref{eq:match}. The red--red and green--green measurements in Figs.~\ref{fig:results-lensless}(b) and \ref{fig:results-lensless}(c) both exhibit a localized correlation peak and an upright correlation image. For the green--green control, the nominal equal-arm distance is $\sim60~\mathrm{cm}$.

For different detection wavelengths, the arm lengths are adjusted to satisfy the wavelength-weighted propagation condition in Eq.~\eqref{eq:match}, i.e. $\lambda_o z_o=\lambda_r z_r$. A red path of $60~\mathrm{cm}$ requires a green path of $71.4~\mathrm{cm}$, consistent with the nominal experimental distances. 
Figure~\ref{fig:results-lensless}(d) shows green illumination of the object with red reference detection, whereas Fig.~\ref{fig:results-lensless}(e) reverses these roles. Both measurements recover the ``GI'' pattern in the coordinates of a wavelength different from that illuminating the object.

These observations are consistent with the same-order kernel in Eq.~\eqref{eq:lenslessimage}: matching the wavelength--distance products cancels the quadratic source phase.
The model therefore predicts an upright image with unit transverse magnification ($m_{same}=+1$), and the unequal physical path lengths do not introduce an additional wavelength-ratio magnification. 
This experiment demonstrates cross-wavelength retrieval of object intensity information through joint detection, with both optical carrier wavelengths unchanged.

\subsubsection{Lens-assisted imaging}

\begin{figure}[htbp]
\centering
\includegraphics[width=0.96\linewidth]{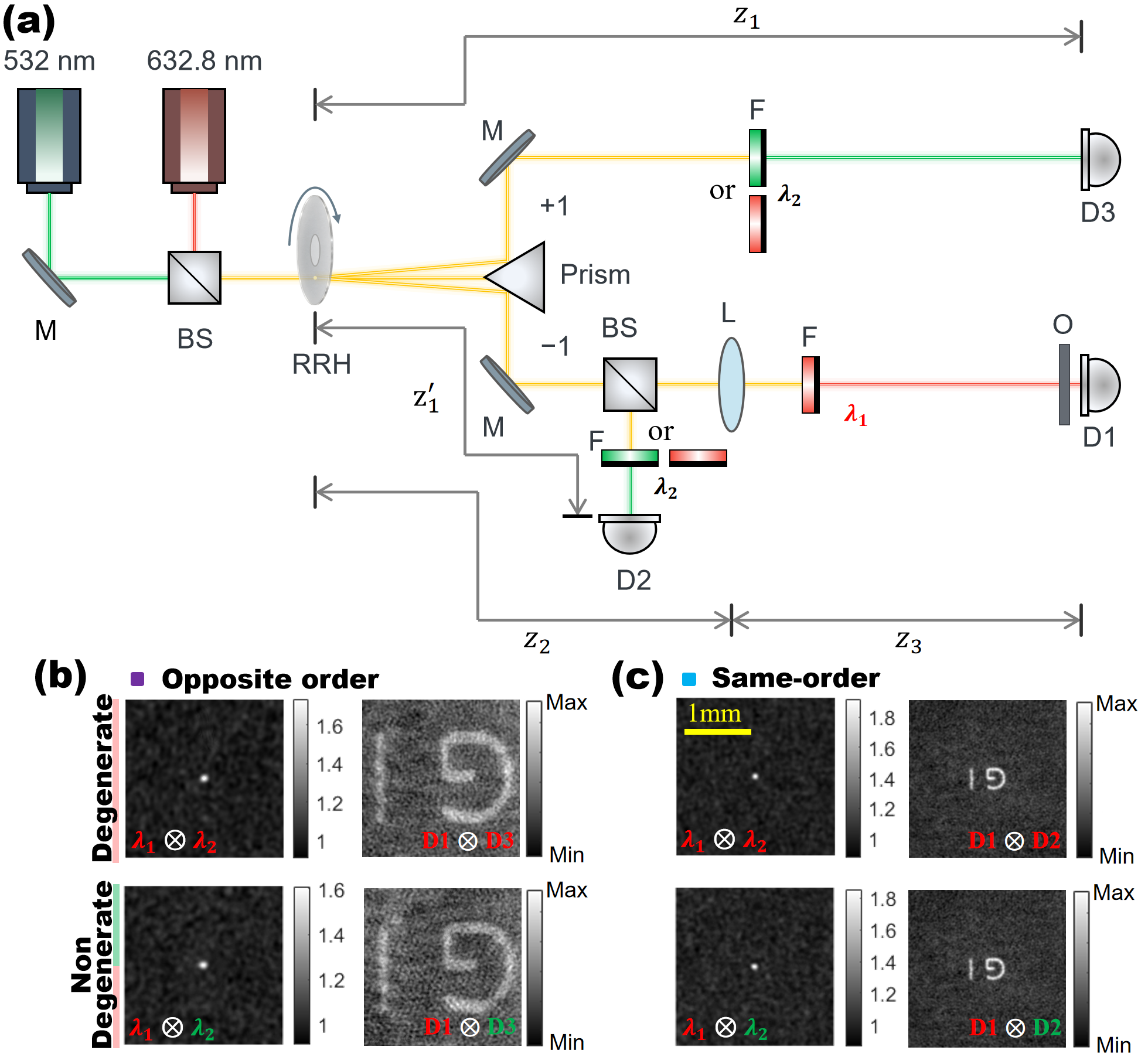}
\caption{Lens-assisted correlation imaging with opposite- and same-order channels. (a) Experimental arrangement. The $-1$ order passes through lens L and illuminates object O at $\lambda_1=632.8~\mathrm{nm}$; D1 supplies the bucket signal. Reference detection at D3 selects the $+1$ order, whereas D2 receives a split copy of the $-1$ order. Alternative reference filters select red or green detection. The distances $z_2$ and $z_3$ denote RRH-to-lens and lens-to-object propagation; $z_1$ and $z_1'$ denote the reference distances to D3 and D2, respectively. (b) Opposite-order D1--D3 results, with enlarged, inverted reconstructions. (c) Same-order D1--D2 results, with reduced, inverted reconstructions. In each panel, the upper row uses degenerate red--red detection and the lower row uses nondegenerate red-object/green-reference detection. Each row shows a normalized point-to-point correlation map $g^{(2)}$ on the left and the corresponding bucket--reference image on the right. The object distances are $z_3=90~\mathrm{cm}$ in (b) and $180~\mathrm{cm}$ in (c), with $z_2=129~\mathrm{cm}$ and $f=60~\mathrm{cm}$ in both configurations. Each reconstruction uses $2\times10^4$ realizations.}
\label{fig:results-lens-assisted}
\end{figure}

Figure~\ref{fig:results-lens-assisted}(a) extends the lensless arrangement by placing an achromatic Fourier lens of focal length $f=60~\mathrm{cm}$ in the object arm, enabling imaging with either diffraction-order pairing.
Following propagation from the RRH, the $\pm1$-order diffracted beams separate in space and are split into two arms by a reflective prism.
The object arm is selected from the $-1$ order and filtered at $\lambda_o=\lambda_1=632.8~\mathrm{nm}$. The ``GI'' mask is placed immediately before D1, which supplies the bucket signal. Two reference channels provide a direct comparison: D3 receives the $+1$ order for opposite-order imaging, whereas D2 receives a split copy of the $-1$ order for same-order imaging. The reconstructions in Figs.~\ref{fig:results-lens-assisted}(b) and \ref{fig:results-lens-assisted}(c) therefore use the D1--D3 and D1--D2 detector pairs, respectively. Exchanging the reference filter between red and green selects degenerate red--red or nondegenerate red-object/green-reference detection, while the object-arm wavelength remains fixed.

The distances in Fig.~\ref{fig:results-lens-assisted}(a) map directly onto Section~\ref{sec:imaging}: $a=z_2$ is the RRH-to-lens distance and $b=z_3$ is the lens-to-object distance. The RRH-to-reference distance is $z_r=z_1'$ for D2 and $z_r=z_1$ for D3. With $\rho=\lambda_r/\lambda_o$, the same-order imaging condition is
\begin{equation}
\frac{1}{z_2-\rho z_1'}+\frac{1}{z_3}=\frac{1}{f},
\label{eq:results-same-summary}
\end{equation}
whereas opposite-order imaging requires
\begin{equation}
\frac{1}{z_2+\rho z_1}+\frac{1}{z_3}=\frac{1}{f}.
\label{eq:results-opp-summary}
\end{equation}
Here $\rho=1$ for red--red detection and $\rho=532/632.8\simeq0.841$ for red--green detection. The minus and plus signs are the two effective propagation branches of Eq.~\eqref{eq:effective}. At their respective conjugate planes, Eq.~\eqref{eq:magnification} gives the object-to-reference magnifications
\begin{equation}
\begin{aligned}
m_{\rm same}&=-\frac{z_2-\rho z_1'}{z_3},\\
m_{\rm opp}&=-\frac{z_2+\rho z_1}{z_3}.
\end{aligned}
\label{eq:results-magnification-summary}
\end{equation}

For the opposite-order measurements in Fig.~\ref{fig:results-lens-assisted}(b), $z_2\approx129~\mathrm{cm}$ and $z_3\approx90~\mathrm{cm}$. Equation~\eqref{eq:results-opp-summary} then requires an effective distance $z_2+\rho z_1=180~\mathrm{cm}$, predicting $z_1=51.0~\mathrm{cm}$ for red reference detection and $60.7~\mathrm{cm}$ for green reference detection. These values are close to the nominal experimental positions of $50$ and $60~\mathrm{cm}$, respectively. Both rows exhibit a localized point-to-point correlation peak and a recognizable bucket--reference reconstruction of the ``GI'' mask. The reconstructed images are enlarged and inverted, consistent with the predicted magnification $m_{\rm opp}=-2$.

For the same-order measurements in Fig.~\ref{fig:results-lens-assisted}(c), the lens-to-object distance is changed to $z_3\approx180~\mathrm{cm}$ while $z_2$ and $f$ remain unchanged. Equation~\eqref{eq:results-same-summary} now requires $z_2-\rho z_1'=90~\mathrm{cm}$, predicting $z_1'=39.0~\mathrm{cm}$ for red reference detection and $46.4~\mathrm{cm}$ for green reference detection. The corresponding experimental positions, approximately $40$ and $47~\mathrm{cm}$, agree within the estimated position uncertainty of $2~\mathrm{cm}$. The localized correlation peaks are accompanied by reduced, inverted reconstructions in both wavelength configurations, consistent with $m_{\rm same}=-0.5$.

\begin{figure}[htbp]
\centering
\includegraphics[width=0.94\linewidth]{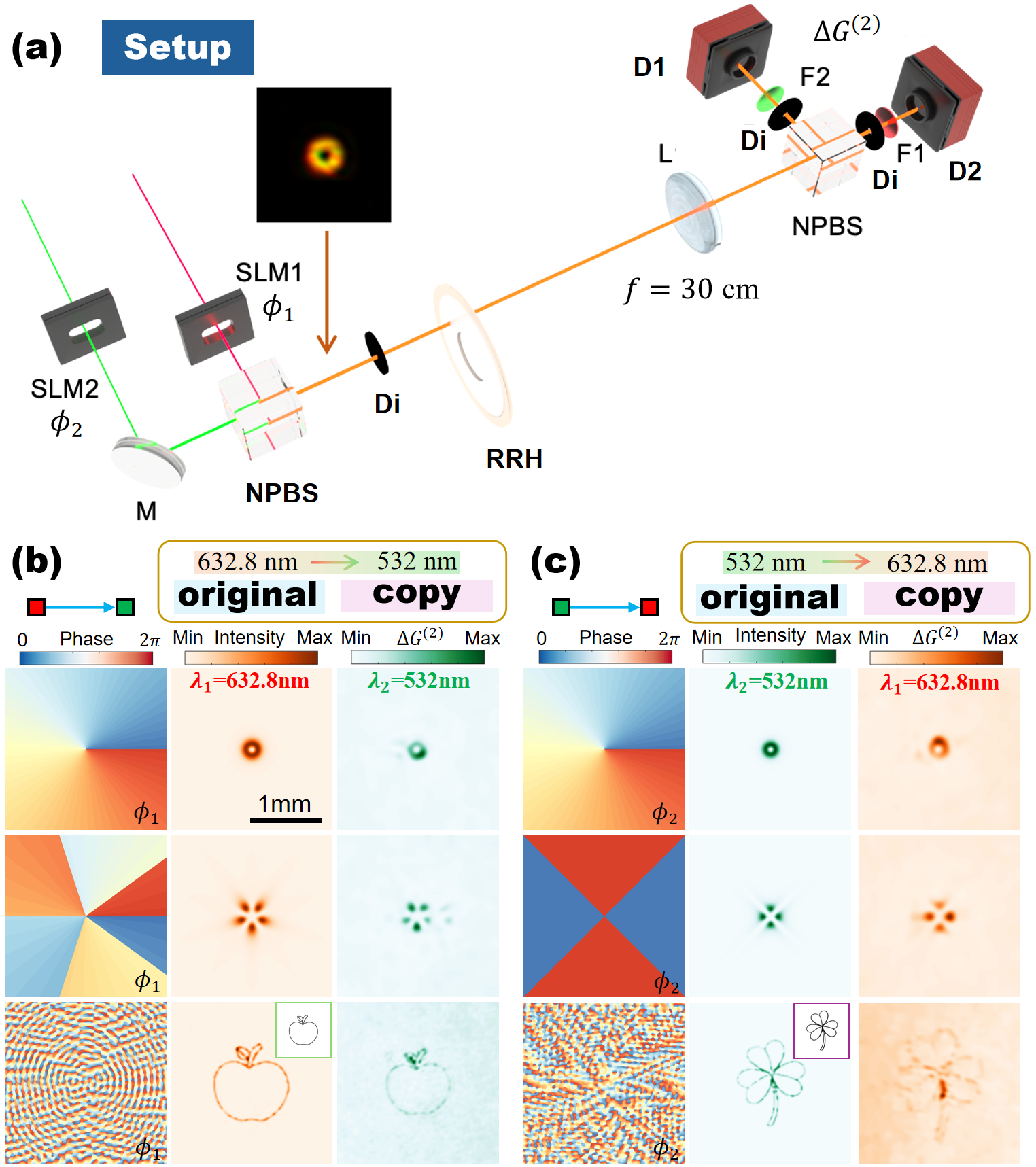}
\caption{Optical replication between $632.8$ and $532~\mathrm{nm}$ from $\pm1$-order diffracted beams. (a) Unified apparatus for replication and spatial information mixing. Transmissive spatial light modulators SLM1 and SLM2 encode the input phase distributions, and the two colors overlap on the rotating random hologram (RRH). L, Fourier lens with $f=30~\mathrm{cm}$; NPBS, non-polarizing beam splitter; Di, diaphragm; M, mirror; F1 and F2, spectral filters; D1 and D2, CMOS-based cameras. (b) Red-to-green replication of a vortex ring, a multi-lobed pattern, and an apple outline. (c) Green-to-red replication of a ring, a multi-lobed pattern, and a flower outline. In each group, the columns show the encoded phase, the associated original intensity, and the connected correlation $\Delta G^{(2)}$ reconstructed in the other wavelength's detection coordinates.
$2\times10^4$ realizations contribute to the image quality of these correlation patterns.
}
\label{fig:results-replication}
\end{figure}

The two panels thus test the signed effective distances of the correlation model through separately matched imaging geometries. The magnification signs refer to the unfolded transverse coordinates used in the theory. In each panel, changing the reference wavelength shifts the required reference plane while preserving the predicted magnification: once the imaging condition is imposed, Eq.~\eqref{eq:magnification} reduces to $m_\beta=-f/(z_3-f)$. The different image sizes in panels (b) and (c) follow from their different object distances, together with the selected order pairing. 
These experimental observations validate the theory of the wavelength-dependent imaging relations by Boyd \textit{et al.}~\cite{chan2009two}, the key to which lies in the establishment of nondegenerate correlations under holographic modulation, thus providing a new resource for cross-wavelength imaging applications.

\subsection{Optical replication across wavelengths}\label{sec:results-replication}

Figure~\ref{fig:results-replication}(a) shows the common experimental platform for optical replication and spatial information mixing. Two transmissive spatial light modulators (SLM1 and SLM2, Daheng Optics, GCI-770104) display holograms encoding the prescribed phase distributions $\phi_1$ and $\phi_2$ to prepare structured inputs at $\lambda_1=632.8~\mathrm{nm}$ and $\lambda_2=532~\mathrm{nm}$, respectively. The beams are combined by a non-polarizing beam splitter (NPBS) and overlap on the RRH. A common lens L of focal length $f=30~\mathrm{cm}$ provides the Fourier readout of Sections~\ref{sec:replication} and \ref{sec:mixing}, before the detection paths are separated by a second NPBS. Diaphragms select the spatial channels ($\pm1$ orders), while spectral filters isolate the required wavelength components at cameras D1 and D2. For the correlation readout, the intensity at a fixed spatial sample in one channel is correlated with the spatially resolved intensity in the other over a large number of realizations. Subtracting the product of the mean intensities gives $\Delta G^{(2)}$ as defined in Eq.~\eqref{eq:covariance}.

For replication, a phase hologram is applied to one input while the other provides a spatially smooth reference. Unlike the preceding imaging experiments, the image information is embedded into the incident coherent beam before illuminating the RRH. Figure~\ref{fig:results-replication}(b) shows replication from the red input to the green detection coordinates over \(2\times10^4\) realizations of spatial averages~\cite{vinu2020ghost,tiwari2023ghost,ye2024hanbury}. 
The original intensity distributions are directly recorded in the Fourier plane without the modulation of the RRH, providing the reference patterns for evaluating the cross-wavelength reconstructions.
The three rows display a vortex-ring pattern, a multi-lobed pattern, and a more complex apple-shaped outline. In each case, the phase distribution and its associated original intensity are compared with the correlation reconstruction in the other color. The reconstructed distributions retain the principal spatial features of the originals. Thus the operation extends beyond the replication of a simple rotationally symmetric mode to spatially nonuniform patterns encoded by phase holograms.

\begin{figure}[htbp]
\centering
\includegraphics[width=1.00\linewidth]{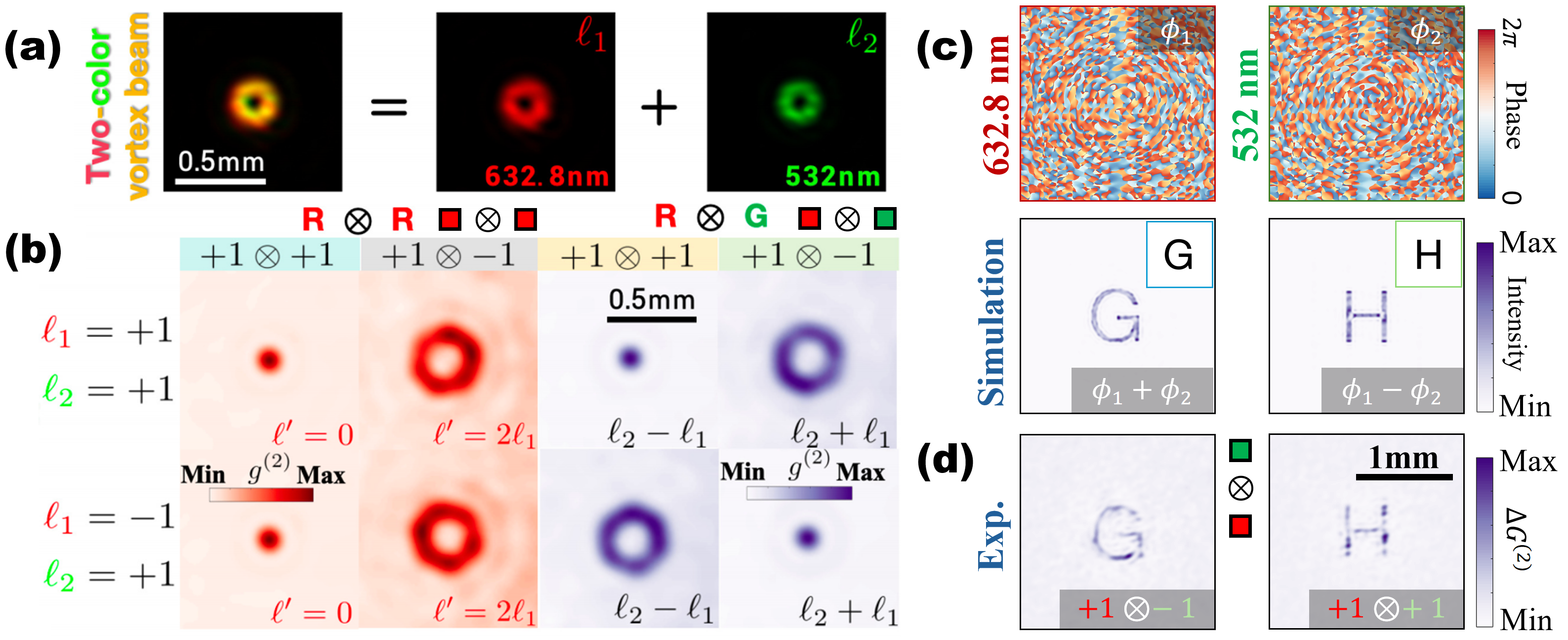}
\caption{Spatial information mixing of vortex inputs and phase-encoded patterns. (a) Overlapped red and green vortex beams with input charges $\ell_1$ and $\ell_2$. (b) Normalized correlation maps $g^{(2)}$ for the degenerate red--red (R$\otimes$R) control and the nondegenerate red--green (R$\otimes$G) measurement. Within each pair, the columns select the same $(+1,+1)$ or opposite $(+1,-1)$ diffraction orders. The upper and lower rows correspond to $(\ell_1,\ell_2)=(+1,+1)$ and $(-1,+1)$, respectively, and the red--red control uses the red input in both channels. (c) Two input phase distributions and simulated propagated intensities of their sum and difference combinations, designed to form ``G'' and ``H''. (d) Experimental observations of connected correlations $\Delta G^{(2)}$ for opposite-order phase addition and same-order phase subtraction. $2\times10^4$ realizations contribute to the image quality of these correlation patterns.
}
\label{fig:results-mixing}
\end{figure}

Reversing the encoding and readout roles gives the green-to-red results in Fig.~\ref{fig:results-replication}(c), where a ring, a multi-lobed distribution, and a flower-shaped outline are recovered. These complementary measurements demonstrate bidirectional retrieval of spatial information between the two wavelengths. The complex contours remain recognizable despite residual background and nonuniform reconstructed brightness. 
As predicted in Eq.~\eqref{eq:replicamap}, since $f_1=f_2=30 \mathrm{cm}$, the magnification is related to $\lambda_1/\lambda_2$.
For equal focal lengths, the copy-to-original transverse scale factor is the ratio of the output to input wavelengths. Red-to-green replication therefore gives a scale factor of approximately 0.84, whereas green-to-red replication gives approximately 1.19. The image orientation is determined by the selected diffraction-order pairing and the registered detector coordinates.
A Gerchberg--Saxton (GS) phase-retrieval procedure~\cite{fienup1982phase} is used in the last row of Figs.~\ref{fig:results-replication}(b)(c) to design phase-only holographic encoding of prescribed propagated intensity patterns.

\subsection{Spatial information mixing of two-color inputs}\label{sec:results-mixing}

When both SLMs in Fig.~\ref{fig:results-replication}(a) encode structured inputs, the same apparatus combines their spatial information through the product or conjugate-product kernels of Section~\ref{sec:mixing}. We first examine two vortex inputs, illustrated in Fig.~\ref{fig:results-mixing}(a). Their azimuthal phases are specified by topological charges $\ell_1$ and $\ell_2$. As predicted by Eq.~\eqref{eq:oam}, opposite-order selection combines the phases through their sum, whereas same-order selection combines them through their difference. 

The red--red measurements provide a degenerate control. Correlating the same orders of the red field cancels its azimuthal phase and produces a central peak, whereas correlating opposite orders gives a ring consistent with the doubled kernel index $2\ell_1$. 
The red--green measurements then combine independently prepared inputs. For $\ell_1=\ell_2=+1$, the same-order correlation has a central maximum, consistent with a zero difference index ($\ell_{\rm same}=\ell_1-\ell_2=0$), while the opposite-order correlation is annular, consistent with a sum index of $\ell_{\rm opp}=\ell_1+\ell_2=+2$. Changing the red input to $\ell_1=-1$ while keeping $\ell_2=+1$ reverses this behavior: the same-order distribution becomes annular and the opposite-order distribution develops a central maximum. These changes agree with the zero versus nonzero azimuthal indices predicted by the two mixing kernels. 

To distinguish the two operations through more general spatial information, we next use the phase distributions in Fig.~\ref{fig:results-mixing}(c) obtained from the phase retrieval algorithm~\cite{fienup1982phase}. 
As shown in the second row of Fig.~\ref{fig:results-mixing}(c), the two phase masks are designed so that Fourier propagation of their complex-envelope product and conjugate product produces the prescribed ``G''-shaped and ``H''-shaped intensity patterns, respectively.
The experimental results of connected correlations in Fig.~\ref{fig:results-mixing}(d) recover ``G'' for the opposite-order diffracted beams and ``H'' for the same-order diffracted beams over $2\times10^4$ realizations. Thus the two encoded inputs jointly specify the reconstruction, and diffraction-order selection determines which target is read out in either wavelength. 
In the sense of Eq.~\eqref{eq:mixing-correspondence}, the ``G'' and ``H'' measurements realize the spatial operations associated with SFG and DFG, respectively, while the optical carrier frequencies remain unchanged. Together with the replication results, these measurements demonstrate that the same correlation platform can retrieve a single encoded input or combine the spatial phase information of two independently prepared structured inputs at different wavelengths.



\section{Discussion}\label{sec4}

The use of $532$ and $632.8~\mathrm{nm}$ in the present experiments represents one spectral pair within a broader operating range. Previous work has demonstrated intensity correlations between $351.1$ and $1064~\mathrm{nm}$ in both same- and opposite-order channels, establishing a link from the ultraviolet to the near-infrared~\cite{ye2026all}. The accessible spectral window is primarily set by the material response of the diffractive modulator, including substrate transparency and the wavelength-dependent phase modulation and diffraction efficiency. 
Different substrates, coatings, and relief depths could therefore be designed for particular, widely separated wavelength pairs. Such devices could provide statistical information-relay channels between visible and non-visible bands, allowing an input prepared or interrogated in one spectral region to be processed and read out in another. The demonstrated ultraviolet--infrared correlations support this prospect, while each new spectral pair would require calibration of its correlation strength and spatial transfer response.

\begin{figure}[t]
\centering
\includegraphics[width=1.00\linewidth]{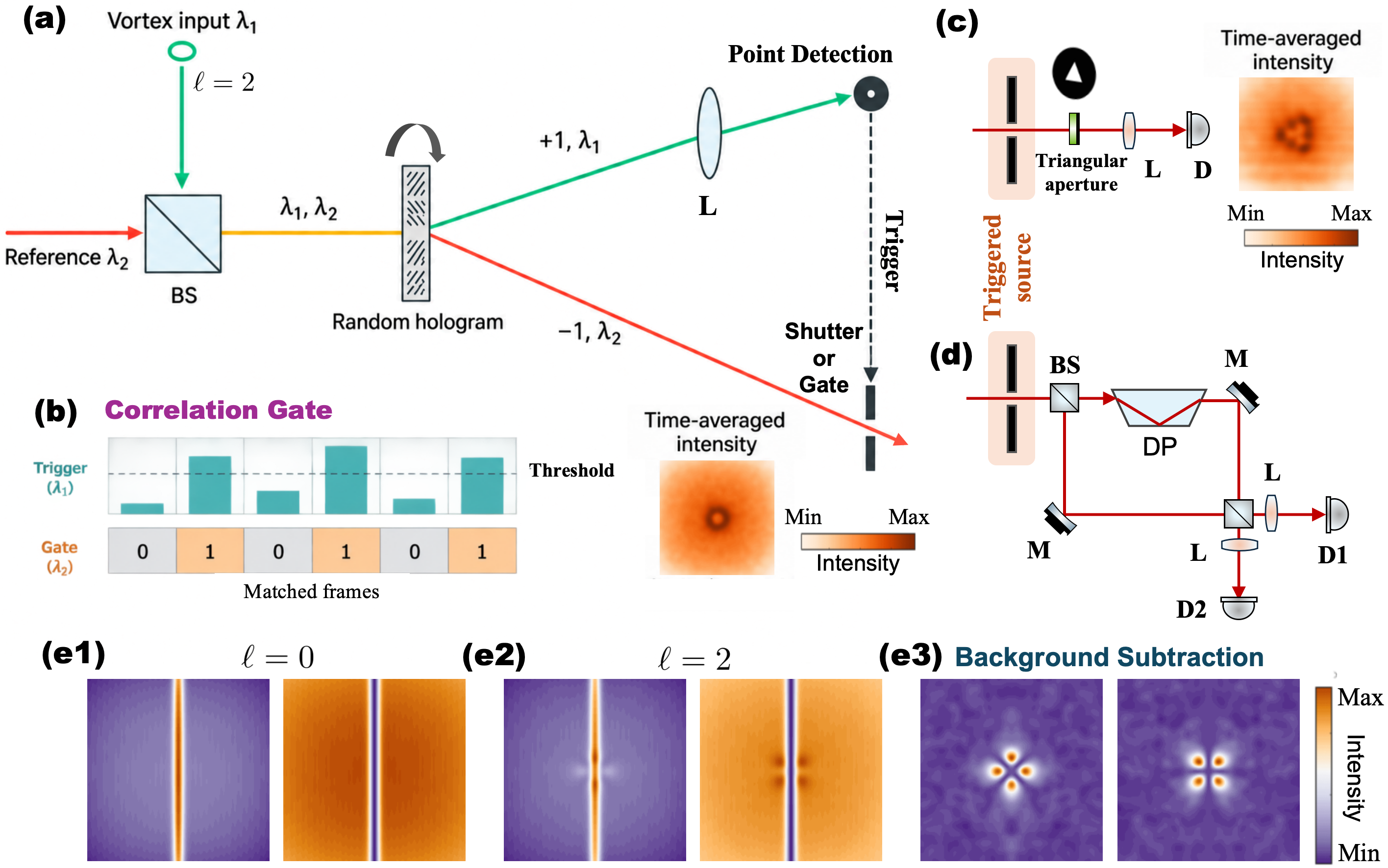}
\caption{Correlation-gated statistical light-field manipulation. (a) Set-up diagram. An input vortex beam ($\ell=2$) at $\lambda_1$ and a reference input at $\lambda_2$ illuminate the RRH. Point detection in the $+1$ channel at $\lambda_1$ controls a global gate in the $-1$ channel at $\lambda_2$. The directly averaged transmitted intensity contains an annular structured contribution plus a background. (b) Threshold selection transmits only the $\lambda_2$ realizations matched to accepted trigger signals. (c) Simulated time-averaged diffraction after the gated field illuminates a triangular aperture. (d) Proposed Mach--Zehnder analysis with relative image inversion between the two arms, implemented using a Dove prism (DP). BS, beam splitter; M, mirror; L, lens; D, detector. (e1,e2) Simulated complementary output intensities for input indices $\ell=0$ and $\ell=2$ at $\lambda_1$, respectively. Background self-interference produces bright and dark vertical features, with additional azimuthal structure for $\ell=2$. (e3) For $\ell=2$, background-subtracted displays reveal complementary four-petal patterns. All detectors here are placed in the Fourier plane (far field). 
}
\label{fig:correlation-gating}
\end{figure}

The preceding experiments retrieve spatial information by evaluating intensity correlations digitally. Figure~\ref{fig:correlation-gating} proposes a further step: using one intensity signal to select the optical realizations transmitted by the other channel. In Fig.~\ref{fig:correlation-gating}(a), a point detector at $\bx_{1,0}$ measures the fluctuating intensity at $\lambda_1$ and drives a spatially uniform gate at $\lambda_2$. Only realizations satisfying a threshold condition are transmitted, as illustrated in Fig.~\ref{fig:correlation-gating}(b). This operation builds on conditional-averaging approaches to thermal-light imaging~\cite{luo2012nonlocal,yang2018probability}, but applies the selection to the propagating light itself. 
Triggered intensified cameras (ICCDs) provide a related conditional-detection strategy~\cite{fickler2013real,aspden2013epr}. The present proposal instead requires an optical switch that selects the propagating field before subsequent diffraction or interference
The information-bearing distribution can then be observed by direct intensity averaging after the gate, without calculating a pixel-by-pixel product of the two detected intensities. 
Such statistical light-field manipulation techniques have also been reported in many unconventional radiation sources~\cite{ceddia2022ghost,ceddia2023universal,kingston2025neutron}.


To make this distinction explicit, define $\mu_1=\avg{I_1(\bx_{1,0})}$ and an ideal binary gate $g=\Theta[I_1(\bx_{1,0})-\nu\mu_1]$, where $\Theta$ is the Heaviside function and $\nu\geq0$ is a dimensionless threshold. For a single resolved speckle mode under the Gaussian assumptions of Section~\ref{sec:source}, the trigger intensity is exponentially distributed and the accepted realization fraction is $p_g=\avg{g}=\mathrm e^{-\nu}$. Conditional averaging gives
\begin{equation}
\begin{aligned}
\avg{I_2(\bx_2)}_g
=\frac{\avg{g I_2(\bx_2)}}{p_g}
=\avg{I_2(\bx_2)}
+\frac{\nu}{\mu_1}\Delta G^{(2)}_\beta(\bx_{1,0},\bx_2),
\end{aligned}
\label{eq:gate-mean}
\end{equation}
where $\beta$ specifies the selected diffraction-order pairing. The physical output averaged over the entire exposure is $\avg{I_{2,\mathrm{out}}}=p_g\avg{I_2}_g$, and the conditional mean in Eq.~\eqref{eq:gate-mean} is normalized by the accepted fraction. Thus the same correlation pattern measured in Sections~\ref{sec:replication} and \ref{sec:mixing} becomes a structured contribution to the transmitted mean intensity. Increasing the threshold enhances this contribution relative to the background, while reducing the accepted fraction exponentially. Spatially integrated triggers or appreciable temporal averaging require the corresponding measured intensity statistics instead of this single-mode expression.

The effect also extends to spatial coherence, which determines how the transmitted light behaves in subsequent optics. Write $W_{2,\beta}(\bx,\bx')=\avg{e_{2,\beta}(\bx)e_{2,\beta}^*(\bx')}$ for the ungated mutual coherence, with $\bx$ and $\bx'$ denoting two points in the second channel. Under the same assumptions,
\begin{equation}
\begin{aligned}
W_{\mathrm{out},\beta}(\bx,\bx')
=p_g\Big[W_{2,\beta}(\bx,\bx')
+\frac{\nu}{\mu_1}K_\beta(\bx)K_\beta^*(\bx')\Big],
\end{aligned}
\label{eq:gate-coherence}
\end{equation}
where $K_{\rm same}(\bx)=\Gamma_{\rm same}^*(\bx_{1,0},\bx)$ and $K_{\rm opp}(\bx)=\Gamma_{\rm opp}(\bx_{1,0},\bx)$ follow the conventions of Eq.~\eqref{eq:kernels}. The additional term describes a single coherent spatial mode enhanced by correlation gating against the existing partially coherent background. With a smooth second input, this mode contains information from $U_1$, for instance, the annular pattern in Fig.~\ref{fig:correlation-gating}(a) corresponding to $\ell=2$. 
With two structured inputs, it contains their product or conjugate-product information through Eq.~\eqref{eq:bilinear}, including the indicated conjugation convention. The transmitted ensemble remains partially coherent, with zero ensemble-mean field. Nevertheless, its modified mutual coherence is an optical resource that can undergo further propagation, diffraction, and interference. Correlation gating therefore provides a route from a numerical correlation readout to statistical replication or mixing encoded in the light transmitted at $\lambda_2$.

The numerical results in Fig.~\ref{fig:correlation-gating} illustrate this possibility. A practical realization could use a Pockels cell with polarizing optics, as demonstrated for nanosecond gating of full images~\cite{bowman2019electro}, or a fast acousto-optic shutter~\cite{schwenger2012high}. Photodetection-driven electro-optic feed-forward with optical-delay compensation has also been demonstrated~\cite{pittman2002demonstration}. For the proposed apparatus, detector response, control latency, and switching time must preserve the correspondence between speckle realizations, either by operating within their decorrelation time or by introducing a suitable optical delay. Aperture, extinction ratio, insertion loss, and spatial uniformity of the gate determine the usable mode bandwidth and contrast. 

Propagation-sensitive tests distinguish the selected structured contribution from an annular intensity distribution alone. In Fig.~\ref{fig:correlation-gating}(c), the triggered or gated source illuminates a triangular aperture and the simulated time-averaged diffraction retains a characteristic triangular structure over a diffuse background. Triangular-aperture diffraction is an established probe of vortex structure~\cite{pinheiro2022observation,de2011measuring}. Here its interpretation follows Eq.~\eqref{eq:gate-coherence}: any subsequent linear optical system acts on both the background coherence and the selected mode $K_\beta$. The resulting diffraction pattern therefore probes information carried by the transmitted ensemble, beyond the intensity profile at the gate output.

Figure~\ref{fig:correlation-gating}(d) provides a complementary coherence test using a Mach--Zehnder interferometer whose two arms produce an image and its mirror-inverted copy. This relative inversion is essential to the angular-coherence readout~\cite{kulkarni2017single}. For a vortex contribution proportional to $\mathrm e^{\mathrm i\ell\varphi}$, the interference term varies as $\cos(2\ell\varphi+\delta)$, where $\delta$ includes the relative phase and orientation of the two arms. The two output ports consequently exhibit complementary patterns with $2|\ell|$ petals; $|\ell|=2$ gives the four-petal structures in Fig.~\ref{fig:correlation-gating}(e3). The unsubtracted maps in Figs.~\ref{fig:correlation-gating}(e1,e2) also contain complementary bright and dark vertical features akin to the observations in \cite{kulkarni2017single}. For matched propagation of the two replicas, relative image inversion is preserved at the observation plane. The vertical features are consistent with self-interference of the propagated background: near the reflection axis, the two replicas coincide, and their mutual-coherence contribution remains nonzero even when correlations between well-separated points are weak. They do not require coherence between independent background fields. Background subtraction makes the vortex contribution more apparent in Fig.~\ref{fig:correlation-gating}(e3), but serves only as a diagnostic visualization. The physical output retains both the structured contribution and the background.

\section{Conclusions}\label{sec5}

To sum up, we have demonstrated a statistical optical platform for transferring and combining spatial information across wavelengths. Experiments at $532$ and $632.8~\mathrm{nm}$ establish nondegenerate correlation imaging, bidirectional replication of simple and complex spatial patterns, and diffraction-order-selective mixing of two structured inputs. The common theoretical description connects these measurements to product and conjugate-product correlation kernels, reproducing the spatial phase operations associated with SFG and DFG while retaining the input optical carrier frequencies. Beyond digital correlation reconstruction, we propose correlation-gated statistical field preparation and support it with simulations of direct intensity averaging, triangular-aperture diffraction, and image-inverting interferometry. Conditional transmission would place the replicated or mixed spatial information in a propagating, partially coherent field with an accompanying background. Extending the modulation materials to other spectral pairs and implementing synchronized optical gating could enable statistical information relays and subsequent spatial processing between visible and non-visible wavelength bands.


\section*{Acknowledgments}
We wish to acknowledge the support of the National Natural Science Foundation of China (12504411,12274037), the Basic and Applied Basic Research Foundation of Guangdong Province (2026A1515011704), the Science and Technology Development Fund from Macau SAR (FDCT) (0105/2023/RIA2), the research outputs of this Open Research Project Programme are funded by the Science and Technology Development Fund of the Macao S.A.R. (472 0002/2024/TFP), and the Fundamental Research Funds for the Central Universities.

\section*{Conflicts of Interest}
The authors declare no conflicts of interest.

\section*{Data Availability Statement}
The data are available from the authors upon reasonable request.

\bibliographystyle{unsrt}
\bibliography{references}

@preamble{ "\providecommand{\BibDOI}[1]{\href{https://doi.org/#1}{\nolinkurl{#1}}}" }

@article{forbes2021structured,
  title={Structured light},
  author={Forbes, Andrew and De Oliveira, Michael and Dennis, Mark R},
  journal={Nature photonics},
  volume={15},
  number={4},
  pages={253--262},
  year={2021},
  publisher={Nature Publishing Group UK London},
  doi={10.1038/s41566-021-00780-4},
  note={DOI: \BibDOI{10.1038/s41566-021-00780-4}}
}

@article{dorrah2022tunable,
  title={Tunable structured light with flat optics},
  author={Dorrah, Ahmed H and Capasso, Federico},
  journal={Science},
  volume={376},
  number={6591},
  pages={eabi6860},
  year={2022},
  publisher={American Association for the Advancement of Science},
  doi={10.1126/science.abi6860},
  note={DOI: \BibDOI{10.1126/science.abi6860}}
}

@article{he2022towards,
  title={Towards higher-dimensional structured light},
  author={He, Chao and Shen, Yijie and Forbes, Andrew},
  journal={Light: Science \& Applications},
  volume={11},
  number={1},
  pages={205},
  year={2022},
  publisher={Nature Publishing Group UK London},
  doi={10.1038/s41377-022-00897-3},
  note={DOI: \BibDOI{10.1038/s41377-022-00897-3}}
}

@article{wang2012terabit,
  title={Terabit free-space data transmission employing orbital angular momentum multiplexing},
  author={Wang, Jian and Yang, Jeng-Yuan and Fazal, Irfan M and Ahmed, Nisar and Yan, Yan and Huang, Hao and Ren, Yongxiong and Yue, Yang and Dolinar, Samuel and Tur, Moshe and others},
  journal={Nature photonics},
  volume={6},
  number={7},
  pages={488--496},
  year={2012},
  publisher={Nature Publishing Group UK London},
  doi={10.1038/nphoton.2012.138},
  note={DOI: \BibDOI{10.1038/nphoton.2012.138}}
}

@article{barh2019parametric,
  title={Parametric upconversion imaging and its applications},
  author={Barh, Ajanta and Rodrigo, Peter John and Meng, Lichun and Pedersen, Christian and Tidemand-Lichtenberg, Peter},
  journal={Advances in Optics and Photonics},
  volume={11},
  number={4},
  pages={952--1019},
  year={2019},
  publisher={Optical Society of America},
  doi={10.1364/AOP.11.000952},
  note={DOI: \BibDOI{10.1364/AOP.11.000952}}
}

@article{yuan2023quantum,
  title={Quantum sensing of microwave electric fields based on Rydberg atoms},
  author={Yuan, Jinpeng and Yang, Wenguang and Jing, Mingyong and Zhang, Hao and Jiao, Yuechun and Li, Weibin and Zhang, Linjie and Xiao, Liantuan and Jia, Suotang},
  journal={Reports on Progress in Physics},
  volume={86},
  number={10},
  pages={106001},
  year={2023},
  publisher={IOP Publishing},
  doi={10.1088/1361-6633/acf22f},
  note={DOI: \BibDOI{10.1088/1361-6633/acf22f}}
}

@article{lei2015massive,
  title={Massive individual orbital angular momentum channels for multiplexing enabled by Dammann gratings},
  author={Lei, Ting and Zhang, Meng and Li, Yuru and Jia, Ping and Liu, Gordon Ning and Xu, Xiaogeng and Li, Zhaohui and Min, Changjun and Lin, Jiao and Yu, Changyuan and others},
  journal={Light: Science \& Applications},
  volume={4},
  number={3},
  pages={e257--e257},
  year={2015},
  publisher={Nature Publishing Group},
  doi={10.1038/lsa.2015.30},
  note={DOI: \BibDOI{10.1038/lsa.2015.30}}
}

@article{radwell2016achromatic,
  title={Achromatic vector vortex beams from a glass cone},
  author={Radwell, Neal and Hawley, RD and G{\"o}tte, JB and Franke-Arnold, Sonja},
  journal={Nature communications},
  volume={7},
  number={1},
  pages={10564},
  year={2016},
  publisher={Nature Publishing Group UK London},
  doi={10.1038/ncomms10564},
  note={DOI: \BibDOI{10.1038/ncomms10564}}
}

@article{buono2022nonlinear,
  title={Nonlinear optics with structured light},
  author={Buono, Wagner Tavares and Forbes, Andrew},
  journal={Opto-Electronic Advances},
  volume={5},
  number={6},
  pages={210174--1},
  year={2022},
  doi={10.29026/oea.2022.210174},
  note={DOI: \BibDOI{10.29026/oea.2022.210174}}
}

@article{zou2022high,
  title={High-capacity free-space optical communications using wavelength-and mode-division-multiplexing in the mid-infrared region},
  author={Zou, Kaiheng and Pang, Kai and Song, Hao and Fan, Jintao and Zhao, Zhe and Song, Haoqian and Zhang, Runzhou and Zhou, Huibin and Minoofar, Amir and Liu, Cong and others},
  journal={Nature Communications},
  volume={13},
  number={1},
  pages={7662},
  year={2022},
  publisher={Nature Publishing Group UK London},
  doi={10.1038/s41467-022-35327-w},
  note={DOI: \BibDOI{10.1038/s41467-022-35327-w}}
}

@article{wang2023coloured,
  title={Coloured vortex beams with incoherent white light illumination},
  author={Wang, Hongtao and Wang, Hao and Ruan, Qifeng and Chan, John You En and Zhang, Wang and Liu, Hailong and Rezaei, Soroosh Daqiqeh and Trisno, Jonathan and Qiu, Cheng-Wei and Gu, Min and others},
  journal={Nature Nanotechnology},
  volume={18},
  number={3},
  pages={264--272},
  year={2023},
  publisher={Nature Publishing Group UK London},
  doi={10.1038/s41565-023-01319-0},
  note={DOI: \BibDOI{10.1038/s41565-023-01319-0}}
}

@article{jang2024wavelength,
  title={Wavelength-multiplexed orbital angular momentum meta-holography},
  author={Jang, Jaehyuck and Moon, Seong-Won and Kim, Joohoon and Mun, Jungho and Maier, Stefan A and Ren, Haoran and Rho, Junsuk},
  journal={PhotoniX},
  volume={5},
  number={1},
  pages={27},
  year={2024},
  publisher={Springer},
  doi={10.1186/s43074-024-00142-7},
  note={DOI: \BibDOI{10.1186/s43074-024-00142-7}}
}

@article{meng2025ultranarrow,
  title={Ultranarrow-linewidth wavelength-vortex metasurface holography},
  author={Meng, Weijia and Fr{\"o}ch, Johannes E and Cheng, Ke and Pi, Dapu and Li, Baoli and Majumdar, Arka and Maier, Stefan A and Ren, Haoran and Gu, Min and Fang, Xinyuan},
  journal={Science Advances},
  volume={11},
  number={12},
  pages={eadt9159},
  year={2025},
  publisher={American Association for the Advancement of Science},
  doi={10.1126/sciadv.adt9159},
  note={DOI: \BibDOI{10.1126/sciadv.adt9159}}
}

@article{ye2025vortex,
  title={Vortex memory effect of light for scattering-assisted massive data transmission},
  author={Ye, Zhiyuan and Ding, Chen-Xin and Zhao, Jilun and Hou, Wanting and Men, Xue-Jiao and Cui, Yi and Liu, Hong-Chao and Wang, Hai-Bo and Xiong, Jun},
  journal={Optica},
  volume={12},
  number={2},
  pages={281--295},
  year={2025},
  publisher={Optica Publishing Group},
  doi={10.1364/OPTICA.544908},
  note={DOI: \BibDOI{10.1364/OPTICA.544908}}
}

@article{ahmed2025customizing,
  title={Customizing multicolored orbital angular momentum combs},
  author={Ahmed, Hammad and Ansari, Muhammad Afnan and Yan, Rong and Chen, Xianzhong},
  journal={Nano Letters},
  volume={25},
  number={13},
  pages={5366},
  year={2025},
  doi={10.1021/acs.nanolett.5c00467},
  note={DOI: \BibDOI{10.1021/acs.nanolett.5c00467}}
}

@article{ji2025full,
  title={Full Color 3D Meta-Holography Based on Achromatic Fresnel Near-Field Diffraction Algorithm},
  author={Ji, Ruo-Nan and Zheng, Xin-Ru and Li, Yi-Long and Xie, Xin and Lin, Fan-Chuan and Liu, Chao and Zheng, Yi-Wei and Yu, Pei-Qi and Li, Xin-Ru and Song, Kun and others},
  journal={Laser \& Photonics Reviews},
  volume={19},
  number={18},
  pages={e00398},
  year={2025},
  publisher={Wiley Online Library},
  doi={10.1002/lpor.202500398},
  note={DOI: \BibDOI{10.1002/lpor.202500398}}
}

@article{han2026wavelength,
  title={Wavelength and Full-Polarization Encrypted Orbital Angular Momentum Meta-Holography},
  author={Han, Yue and Xun, Mingna and He, Yuqing and Ye, Sheng and Wan, Weiping and Li, Yan},
  journal={Laser \& Photonics Reviews},
  pages={e71293},
  year={2026},
  publisher={Wiley Online Library},
  volume={20},
  number={17},
  doi={10.1002/lpor.71293},
  note={DOI: \BibDOI{10.1002/lpor.71293}}
}

@article{singh2026progress,
  title={Progress in structured light with nonlinear optics},
  author={Singh, Sachleen and Forbes, Andrew},
  journal={Prog. Electromagn. Res},
  volume={185},
  pages={1--16},
  year={2026},
  doi={10.2528/PIER25110804},
  note={DOI: \BibDOI{10.2528/PIER25110804}}
}

@article{li2015sum,
  title={Sum frequency generation with two orbital angular momentum carrying laser beams},
  author={Li, Yan and Zhou, Zhi-Yuan and Ding, Dong-Sheng and Shi, Bao-Sen},
  journal={Journal of the Optical Society of America B},
  volume={32},
  number={3},
  pages={407--411},
  year={2015},
  publisher={Optical Society of America},
  doi={10.1364/JOSAB.32.000407},
  note={DOI: \BibDOI{10.1364/JOSAB.32.000407}}
}

@article{steinlechner2016frequency,
  title={Frequency conversion of structured light},
  author={Steinlechner, Fabian and Hermosa, Nathaniel and Pruneri, Valerio and Torres, Juan P},
  journal={Scientific reports},
  volume={6},
  number={1},
  pages={21390},
  year={2016},
  publisher={Nature Publishing Group UK London},
  doi={10.1038/srep21390},
  note={DOI: \BibDOI{10.1038/srep21390}}
}

@article{pereira2017orbital,
  title={Orbital-angular-momentum mixing in type-II second-harmonic generation},
  author={Pereira, Leonardo J and Buono, Wagner T and Tasca, Daniel S and Dechoum, Kaled and Khoury, Antonio Z},
  journal={Physical Review A},
  volume={96},
  number={5},
  pages={053856},
  year={2017},
  publisher={APS},
  doi={10.1103/PhysRevA.96.053856},
  note={DOI: \BibDOI{10.1103/PhysRevA.96.053856}}
}

@article{pinheiro2022observation,
  title={Observation of a triangular-lattice pattern in nonlinear wave mixing with optical vortices},
  author={Pinheiro da Silva, B and Dos Santos, GH and De Oliveira, AG and Rubiano da Silva, N and Buono, WT and Gomes, RdM and Soares, WC and Jesus-Silva, AJ and Fonseca, EJdS and Souto Ribeiro, PH and others},
  journal={Optica},
  volume={9},
  number={8},
  pages={908--912},
  year={2022},
  publisher={Optica Publishing Group},
  doi={10.1364/OPTICA.459812},
  note={DOI: \BibDOI{10.1364/OPTICA.459812}}
}

@article{xu2023orthogonal,
  title={Orthogonal spatial coding with stimulated parametric down-conversion},
  author={Xu, Yang and Tang, Sirui and Nicholas Black, A and Boyd, Robert W},
  journal={Optics Express},
  volume={31},
  number={25},
  pages={42723--42729},
  year={2023},
  publisher={Optica Publishing Group},
  doi={10.1364/OE.506383},
  note={DOI: \BibDOI{10.1364/OE.506383}}
}

@article{yuan2024high,
  title={High-Fidelity Frequency Converter in High-Dimensional Spaces},
  author={Yuan, Jinpeng and Wang, Xuewen and Chen, Gang and Wang, Lirong and Xiao, Liantuan and Jia, Suotang},
  journal={Laser \& Photonics Reviews},
  volume={18},
  number={11},
  pages={2400368},
  year={2024},
  publisher={Wiley Online Library},
  doi={10.1002/lpor.202400368},
  note={DOI: \BibDOI{10.1002/lpor.202400368}}
}

@article{sanchez2025nonlinear,
  title={Nonlinear modal decomposition of structured light},
  author={S{\'a}nchez-Montes, Adriana R and Singh, Sachleen and M{\'a}rquez, Andr{\'e}s and Franc{\'e}s, Jorge and Forbes, Andrew and Dudley, Angela},
  journal={Optics Express},
  volume={33},
  number={19},
  pages={41261--41270},
  year={2025},
  publisher={Optica Publishing Group},
  doi={10.1364/OE.567825},
  note={DOI: \BibDOI{10.1364/OE.567825}}
}

@article{aguilar2025tailoring,
  title={Tailoring spatial modes produced by stimulated parametric down-conversion},
  author={Aguilar-Cardoso, AA and Li, C and Luck, TJB and Ferrer-Garcia, MF and Upham, J and Lundeen, JS and Boyd, RW},
  journal={Physical Review A},
  volume={112},
  number={4},
  pages={043541},
  year={2025},
  publisher={APS},
  doi={10.1103/g22l-cgr1},
  note={DOI: \BibDOI{10.1103/g22l-cgr1}}
}

@article{li2025mid,
  title={Mid-infrared nonlinear pinhole imaging},
  author={Li, Yanan and Huang, Kun and Fang, Jianan and Wei, Zhuohang and Zeng, Heping},
  journal={Optica},
  volume={12},
  number={9},
  pages={1478--1485},
  year={2025},
  publisher={Optica Publishing Group},
  doi={10.1364/OPTICA.566042},
  note={DOI: \BibDOI{10.1364/OPTICA.566042}}
}

@article{qiu2018spiral,
  title={Spiral phase contrast imaging in nonlinear optics: seeing phase objects using invisible illumination},
  author={Qiu, Xiaodong and Li, Fangshu and Zhang, Wuhong and Zhu, Zhihan and Chen, Lixiang},
  journal={Optica},
  volume={5},
  number={2},
  pages={208--212},
  year={2018},
  publisher={Optical Society of America},
  doi={10.1364/OPTICA.5.000208},
  note={DOI: \BibDOI{10.1364/OPTICA.5.000208}}
}

@article{hong2020second,
  title={Second harmonic generation based joint transform correlator for human face and QR code recognitions},
  author={Hong, Ling and Lin, Fei and Qiu, Xiaodong and Chen, Lixiang},
  journal={Applied Physics Letters},
  volume={116},
  number={23},
  year={2020},
  publisher={AIP Publishing},
  pages={231101},
  doi={10.1063/5.0001301},
  note={DOI: \BibDOI{10.1063/5.0001301}}
}

@article{singh2024light,
  title={Light correcting light with nonlinear optics},
  author={Singh, Sachleen and Sephton, Bereneice and Tavares Buono, Wagner and D’Ambrosio, Vincenzo and Konrad, Thomas and Forbes, Andrew},
  journal={Advanced Photonics},
  volume={6},
  number={2},
  pages={026003--026003},
  year={2024},
  publisher={Society of Photo-Optical Instrumentation Engineers},
  doi={10.1117/1.AP.6.2.026003},
  note={DOI: \BibDOI{10.1117/1.AP.6.2.026003}}
}

@article{chen2025observation,
  title={Observation of spatial differentiation in structured nonlinear optics},
  author={Chen, Yi-Zhe and Yan, Ding and Qi, Tong and Wang, Xiang-Wei and Gao, Wei},
  journal={Laser \& Photonics Reviews},
  volume={19},
  number={20},
  pages={e00595},
  year={2025},
  publisher={Wiley Online Library},
  doi={10.1002/lpor.202500595},
  note={DOI: \BibDOI{10.1002/lpor.202500595}}
}

@article{pittman1995optical,
  title={Optical imaging by means of two-photon quantum entanglement},
  author={Pittman, Todd B and Shih, Yanhua H and Strekalov, Dmitry V and Sergienko, Alexander V},
  journal={Physical Review A},
  volume={52},
  number={5},
  pages={R3429},
  year={1995},
  publisher={APS},
  doi={10.1103/PhysRevA.52.R3429},
  note={DOI: \BibDOI{10.1103/PhysRevA.52.R3429}}
}

@article{aspden2015photon,
  title={Photon-sparse microscopy: visible light imaging using infrared illumination},
  author={Aspden, Reuben S and Gemmell, Nathan R and Morris, Peter A and Tasca, Daniel S and Mertens, Lena and Tanner, Michael G and Kirkwood, Robert A and Ruggeri, Alessandro and Tosi, Alberto and Boyd, Robert W and others},
  journal={Optica},
  volume={2},
  number={12},
  pages={1049--1052},
  year={2015},
  publisher={Optical Society of America},
  doi={10.1364/OPTICA.2.001049},
  note={DOI: \BibDOI{10.1364/OPTICA.2.001049}}
}

@article{ma2025quantum,
  title={Quantum imaging using spatially entangled photon pairs from a nonlinear metasurface},
  author={Ma, Jinyong and Ren, Jinliang and Zhang, Jihua and Meng, Jiajun and McManus-Barrett, Caitlin and Crozier, Kenneth B and Sukhorukov, Andrey A},
  journal={ELight},
  volume={5},
  number={1},
  pages={2},
  year={2025},
  publisher={Springer},
  doi={10.1186/s43593-024-00080-8},
  note={DOI: \BibDOI{10.1186/s43593-024-00080-8}}
}

@article{ferri2005high,
  title={High-resolution ghost image and ghost diffraction experiments with thermal light},
  author={Ferri, Fabio and Magatti, Davide and Gatti, Alessandra and Bache, Morten and Brambilla, Enrico and Lugiato, Luigi A},
  journal={Physical review letters},
  volume={94},
  number={18},
  pages={183602},
  year={2005},
  publisher={APS},
  doi={10.1103/PhysRevLett.94.183602},
  note={DOI: \BibDOI{10.1103/PhysRevLett.94.183602}}
}

@article{ye2025random,
  title={Random Holography: Generating EPR-Like Correlation with Thermal Photons},
  author={Ye, Zhiyuan and Hou, Wanting and Ding, Chen-Xin and Men, Xue-Jiao and He, Run-Jie and Zhao, Jilun and Wang, Hai-Bo and Xiong, Jun and Wang, Kaige},
  journal={Laser \& Photonics Reviews},
  volume={19},
  number={6},
  pages={2401610},
  year={2025},
  publisher={Wiley Online Library},
  doi={10.1002/lpor.202401610},
  note={DOI: \BibDOI{10.1002/lpor.202401610}}
}

@article{ye2026all,
  title={All-purpose spatial correlations in holographic thermal light},
  author={Ye, Zhiyuan and Hou, Wanting and He, Run-Jie and Cui, Yi and Liu, Hong-Chao and Wang, Hai-Bo and Xiong, Jun and Wang, Kaige},
  journal={Optica},
  volume={13},
  number={4},
  pages={566--576},
  year={2026},
  publisher={Optica Publishing Group},
  doi={10.1364/OPTICA.586022},
  note={DOI: \BibDOI{10.1364/OPTICA.586022}}
}

@article{hou2025manipulating,
  title={Manipulating classical triple correlations for optical information processing and metrology},
  author={Hou, Wanting and He, Run-Jie and Ye, Zhiyuan and Men, Xue-Jiao and Ding, Chen-Xin and Liu, Hong-Chao and Wang, Hai-Bo and Xiong, Jun},
  journal={Photonics Research},
  volume={13},
  number={8},
  pages={2073--2087},
  year={2025},
  publisher={Chinese Laser Press and Optica Publishing Group},
  doi={10.1364/PRJ.559681},
  note={DOI: \BibDOI{10.1364/PRJ.559681}}
}

@article{kulkarni2017single,
  title={Single-shot measurement of the orbital-angular-momentum spectrum of light},
  author={Kulkarni, Girish and Sahu, Rishabh and Maga{\~n}a-Loaiza, Omar S and Boyd, Robert W and Jha, Anand K},
  journal={Nature communications},
  volume={8},
  number={1},
  pages={1054},
  year={2017},
  publisher={Nature Publishing Group UK London},
  doi={10.1038/s41467-017-01215-x},
  note={DOI: \BibDOI{10.1038/s41467-017-01215-x}}
}

@article{pittman2002demonstration,
  title={Demonstration of feed-forward control for linear optics quantum computation},
  author={Pittman, TB and Jacobs, BC and Franson, JD},
  journal={Physical Review A},
  volume={66},
  number={5},
  pages={052305},
  year={2002},
  publisher={APS},
  doi={10.1103/PhysRevA.66.052305},
  note={DOI: \BibDOI{10.1103/PhysRevA.66.052305}}
}

@article{schwenger2012high,
  title={High-speed acousto-optic shutter with no optical frequency shift},
  author={Schwenger, WJ and Higbie, James M},
  journal={Review of Scientific Instruments},
  volume={83},
  number={8},
  year={2012},
  publisher={AIP Publishing},
  pages={083110},
  doi={10.1063/1.4746292},
  note={DOI: \BibDOI{10.1063/1.4746292}}
}

@article{bowman2019electro,
  title={Electro-optic imaging enables efficient wide-field fluorescence lifetime microscopy},
  author={Bowman, Adam J and Klopfer, Brannon B and Juffmann, Thomas and Kasevich, Mark A},
  journal={Nature communications},
  volume={10},
  number={1},
  pages={4561},
  year={2019},
  publisher={Nature Publishing Group UK London},
  doi={10.1038/s41467-019-12535-5},
  note={DOI: \BibDOI{10.1038/s41467-019-12535-5}}
}

@article{chan2009two,
  title={Two-color ghost imaging},
  author={Chan, Kam Wai Clifford and O’Sullivan, Malcolm N and Boyd, Robert W},
  journal={Physical Review A—Atomic, Molecular, and Optical Physics},
  volume={79},
  number={3},
  pages={033808},
  year={2009},
  publisher={APS},
  doi={10.1103/PhysRevA.79.033808},
  note={DOI: \BibDOI{10.1103/PhysRevA.79.033808}}
}

@article{ye2020pseudo,
  title={Pseudo-thermal ghost imaging with “learned” wavelength conversion},
  author={Ye, Zhiyuan and Sheng, Dian and Hao, Zhendong and Wang, Hai-Bo and Xiong, Jun and Wang, Xia and Jin, Weiqi},
  journal={Applied Physics Letters},
  volume={117},
  number={9},
  year={2020},
  publisher={AIP Publishing},
  pages={091103},
  doi={10.1063/5.0020855},
  note={DOI: \BibDOI{10.1063/5.0020855}}
}

@article{deng2018single,
  title={Single-shot thermal ghost imaging using wavelength-division multiplexing},
  author={Deng, Chao and Suo, Jinli and Wang, Yuwang and Zhang, Zhili and Dai, Qionghai},
  journal={Applied Physics Letters},
  volume={112},
  number={5},
  year={2018},
  publisher={AIP Publishing},
  pages={051107},
  doi={10.1063/1.5001750},
  note={DOI: \BibDOI{10.1063/1.5001750}}
}

@article{xiong2025polychromatic,
  title={Polychromatic ghost interference using supercontinuum pseudothermal light},
  author={Xiong, Jiahao and He, Run-Jie and Ye, Zhiyuan and Liu, Hong-Chao},
  journal={Optics Letters},
  volume={50},
  number={18},
  pages={5586--5589},
  year={2025},
  publisher={Optica Publishing Group},
  doi={10.1364/OL.565075},
  note={DOI: \BibDOI{10.1364/OL.565075}}
}

@article{yu2016ghost,
  title={Ghost imaging with different frequencies through non-degenerated four-wave mixing},
  author={Yu, Ya and Wang, Chengyuan and Liu, Jun and Wang, Jinwen and Cao, Mingtao and Wei, Dong and Gao, Hong and Li, Fuli},
  journal={Optics Express},
  volume={24},
  number={16},
  pages={18290--18296},
  year={2016},
  publisher={Optical Society of America},
  doi={10.1364/OE.24.018290},
  note={DOI: \BibDOI{10.1364/OE.24.018290}}
}

@article{shu2026seeing,
  title={Seeing infrared objects with a wavelength non-degenerate thermal light source},
  author={Shu, Weiqian and Chen, Zilin and Zhang, Wuhong and Boyd, Robert and Chen, Lixiang},
  journal={Optica},
  volume={13},
  number={7},
  pages={1205--1211},
  year={2026},
  publisher={Optica Publishing Group},
  doi={10.1364/OPTICA.595564},
  note={DOI: \BibDOI{10.1364/OPTICA.595564}}
}

@article{fickler2013real,
  title={Real-time imaging of quantum entanglement},
  author={Fickler, Robert and Krenn, Mario and Lapkiewicz, Radek and Ramelow, Sven and Zeilinger, Anton},
  journal={Scientific reports},
  volume={3},
  number={1},
  pages={1914},
  year={2013},
  publisher={Nature Publishing Group UK London},
  doi={10.1038/srep01914},
  note={DOI: \BibDOI{10.1038/srep01914}}
}

@article{aspden2013epr,
  title={EPR-based ghost imaging using a single-photon-sensitive camera},
  author={Aspden, Reuben S and Tasca, Daniel S and Boyd, Robert W and Padgett, Miles J},
  journal={New Journal of Physics},
  volume={15},
  number={7},
  pages={073032},
  year={2013},
  publisher={IOP Publishing},
  doi={10.1088/1367-2630/15/7/073032},
  note={DOI: \BibDOI{10.1088/1367-2630/15/7/073032}}
}

@article{ceddia2022ghost,
  title={Ghost projection},
  author={Ceddia, David and Paganin, David M},
  journal={Physical Review A},
  volume={105},
  number={1},
  pages={013512},
  year={2022},
  publisher={APS},
  doi={10.1103/PhysRevA.105.013512},
  note={DOI: \BibDOI{10.1103/PhysRevA.105.013512}}
}

@article{ceddia2023universal,
  title={Universal mask for hard x rays},
  author={Ceddia, David and Aminzadeh, Alaleh and Cook, Philip K and Pelliccia, Daniele and Kingston, Andrew M and Paganin, David M},
  journal={Optica},
  volume={10},
  number={8},
  pages={1067--1073},
  year={2023},
  publisher={Optica Publishing Group},
  doi={10.1364/OPTICA.490006},
  note={DOI: \BibDOI{10.1364/OPTICA.490006}}
}

@article{kingston2025neutron,
  title={Neutron beam shaping by ghost projection},
  author={Kingston, Andrew M and Aminzadeh, Alaleh and Roberts, Lindon and Brown, Jeremy MC and Salvemini, Filomena and Bevitt, Joseph J and Garbe, Ulf and Paganin, David M},
  journal={Physical Review A},
  volume={111},
  number={4},
  pages={043520},
  year={2025},
  publisher={APS},
  doi={10.1103/PhysRevA.111.043520},
  note={DOI: \BibDOI{10.1103/PhysRevA.111.043520}}
}

@article{picinbono1997second,
  title={Second-order statistics of complex signals},
  author={Picinbono, Bernard and Bondon, Pascal},
  journal={IEEE Transactions on Signal Processing},
  volume={45},
  number={2},
  pages={411--420},
  year={1997},
  publisher={IEEE},
  doi={10.1109/78.554305},
  note={DOI: \BibDOI{10.1109/78.554305}}
}

@article{cui2026correlation,
  title={Correlation visibility and generalized Siegert relation for random light beams},
  author={Cui, Yi and Hou, Wanting and Xiong, Jun and Ye, Zhiyuan},
  journal={Chinese Physics B},
  year={2026},
  doi={10.1088/1674-1056/ae9fc1},
  note={DOI: \BibDOI{10.1088/1674-1056/ae9fc1}}
}

@article{yao2011orbital,
  title={Orbital angular momentum: origins, behavior and applications},
  author={Yao, Alison M and Padgett, Miles J},
  journal={Advances in optics and photonics},
  volume={3},
  number={2},
  pages={161--204},
  year={2011},
  publisher={Optical Society of America},
  doi={10.1364/AOP.3.000161},
  note={DOI: \BibDOI{10.1364/AOP.3.000161}}
}

@article{cao2005geometrical,
  title={Geometrical optics in correlated imaging systems},
  author={Cao, De-Zhong and Xiong, Jun and Wang, Kaige},
  journal={Physical Review A—Atomic, Molecular, and Optical Physics},
  volume={71},
  number={1},
  pages={013801},
  year={2005},
  publisher={APS},
  doi={10.1103/PhysRevA.71.013801},
  note={DOI: \BibDOI{10.1103/PhysRevA.71.013801}}
}

@article{picinbono1994circularity,
  title={On circularity},
  author={Picinbono, Bernard},
  journal={IEEE Transactions on signal processing},
  volume={42},
  number={12},
  pages={3473--3482},
  year={1994},
  publisher={IEEE},
  doi={10.1109/78.340781},
  note={DOI: \BibDOI{10.1109/78.340781}}
}

@article{ye2024hanbury,
  title={Hanbury Brown and Twiss-type optical secret sharing},
  author={Ye, Zhiyuan and Ding, Chen-Xin and Men, Xue-Jiao and Wang, Hai-Bo and Xiong, Jun},
  journal={Optics Express},
  volume={32},
  number={25},
  pages={45142--45157},
  year={2024},
  publisher={Optica Publishing Group},
  doi={10.1364/OE.541294},
  note={DOI: \BibDOI{10.1364/OE.541294}}
}

@article{vinu2020ghost,
  title={Ghost diffraction holographic microscopy},
  author={Vinu, RV and Chen, Ziyang and Singh, Rakesh Kumar and Pu, Jixiong},
  journal={Optica},
  volume={7},
  number={12},
  pages={1697--1704},
  year={2020},
  publisher={Optical Society of America},
  doi={10.1364/OPTICA.409886},
  note={DOI: \BibDOI{10.1364/OPTICA.409886}}
}

@article{tiwari2023ghost,
  title={Ghost diffraction: A spatial statistical approach},
  author={Tiwari, Vipin and Bisht, Nandan S and Das, Bhargab and Singh, Rakesh Kumar and others},
  journal={Optics Communications},
  volume={528},
  pages={129002},
  year={2023},
  publisher={Elsevier},
  doi={10.1016/j.optcom.2022.129002},
  note={DOI: \BibDOI{10.1016/j.optcom.2022.129002}}
}

@article{fienup1982phase,
  title={Phase retrieval algorithms: a comparison},
  author={Fienup, James R},
  journal={Applied optics},
  volume={21},
  number={15},
  pages={2758--2769},
  year={1982},
  publisher={Optical Society of America},
  doi={10.1364/AO.21.002758},
  note={DOI: \BibDOI{10.1364/AO.21.002758}}
}

@article{de2011measuring,
  title={Measuring vortex charge with a triangular aperture},
  author={de Araujo, Lu{\'\i}s EE and Anderson, Matthew E},
  journal={Optics letters},
  volume={36},
  number={6},
  pages={787--789},
  year={2011},
  publisher={Optical Society of America},
  doi={10.1364/OL.36.000787},
  note={DOI: \BibDOI{10.1364/OL.36.000787}}
}

@article{yang2018probability,
  title={Probability theory in conditional-averaging ghost imaging with thermal light},
  author={Yang, Huan and Wu, Shuang and Wang, Hai-Bo and Cao, De-Zhong and Zhang, Su-Heng and Xiong, Jun and Wang, Kaige},
  journal={Physical Review A},
  volume={98},
  number={5},
  pages={053853},
  year={2018},
  publisher={APS},
  doi={10.1103/PhysRevA.98.053853},
  note={DOI: \BibDOI{10.1103/PhysRevA.98.053853}}
}

@article{luo2012nonlocal,
  title={Nonlocal imaging by conditional averaging of random reference measurements},
  author={Luo, Kai-Hong and Huang, Bo-Qiang and Zheng, Wei-Mou and Wu, Ling-An},
  journal={Chinese Physics Letters},
  volume={29},
  number={7},
  pages={074216},
  year={2012},
  doi={10.1088/0256-307X/29/7/074216},
  note={DOI: \BibDOI{10.1088/0256-307X/29/7/074216}}
}




\end{document}